\documentclass[aps,prb,twocolumn,shortbibliography,superscriptaddress]{revtex4-1}
\usepackage{epsfig}
\usepackage{epstopdf}
\usepackage{amsmath}
\usepackage{amsfonts}
\usepackage{amssymb}
\usepackage{hyperref}
\usepackage{bm}
\usepackage{makecell}
\usepackage{rotating}
\usepackage{hyperref}

\usepackage{graphicx}% Include figure files
\usepackage{dcolumn}% Align table columns on decimal point
\usepackage{bm}% bold math
\usepackage{color}

\usepackage{tikz,xcolor,hyperref}

\definecolor{lime}{HTML}{A6CE39}
\DeclareRobustCommand{\orcidicon}{%
	\begin{tikzpicture}
	\draw[lime, fill=lime] (0,0)
	circle [radius=0.16]
	node[white] {{\fontfamily{qag}\selectfont \tiny ID}};
	\draw[white, fill=white] (-0.0625,0.095)
	circle [radius=0.007];
	\end{tikzpicture}
	\hspace{-2mm}
}

\foreach \x in {A, ..., Z}{%
	\expandafter\xdef\csname orcid\x\endcsname{\noexpand\href{https://orcid.org/\csname orcidauthor\x\endcsname}{\noexpand\orcidicon}}
}

\begin{document}

\title{Altermagnetic surface states: towards the observation and utilization of \\ altermagnetism in thin films, interfaces and topological materials}

\author{Raghottam M Sattigeri\orcidC}
\email{rsattigeri@magtop.ifpan.edu.pl}
\affiliation{International Research Centre Magtop, Institute of Physics, Polish Academy of Sciences, Aleja Lotnik\'ow 32/46, PL-02668 Warsaw, Poland}

\author{Giuseppe Cuono\orcidD}
\email{gcuono@magtop.ifpan.edu.pl}
\affiliation{International Research Centre Magtop, Institute of Physics, Polish Academy of Sciences, Aleja Lotnik\'ow 32/46, PL-02668 Warsaw, Poland}

%\author{Tomasz Dietl\orcidB}
%\affiliation{International Research Centre Magtop, Institute of Physics, Polish Academy of Sciences, Aleja Lotnik\'ow 32/46, PL-02668 Warsaw, Poland}

\author{Carmine Autieri\orcidA}
\email{autieri@magtop.ifpan.edu.pl}
\affiliation{International Research Centre Magtop, Institute of Physics, Polish Academy of Sciences, Aleja Lotnik\'ow 32/46, PL-02668 Warsaw, Poland}

\date{\today}
\begin{abstract}
The altermagnetism influences the electronic states allowing the presence of non-relativistic spin-splittings. Since altermagnetic spin-splitting is present along specific $k$-paths of the 3D Brillouin zone, we expect that the altermagnetic surface states will be present on specific surface orientations.
We unveil the properties of the altermagnetic surface states considering three representative space groups: tetragonal, orthorhombic and hexagonal.
We calculate the 2D projected Brillouin zone from the 3D Brillouin zone. We study the surfaces with their respective 2D Brillouin zones establishing where the spin-splittings with opposite sign merge annihilating the altermagnetic properties and on which surfaces the altermagnetism is preserved.
Looking at the three principal surface orientations, we find that for several cases two surfaces are blind to the altermagnetism, while the altermagnetism survives for one surface orientation. Which surface preserves the altermagnetism depends also on the magnetic order. We qualitatively show that an electric field orthogonal to the blind surface can activate the altermagnetism.
Our results predict which surfaces to cleave in order to preserve altermagnetism in surfaces or interfaces and this paves the way to observe non-relativistic altermagnetic spin-splitting in thin films via spin-resolved ARPES and to interface the altermagnetism with other collective modes. We open future perspectives for the study of altermagnetic effects on the trivial and topological surface states.
\end{abstract}


\maketitle

\section{Introduction}

% interfaces and surfaces
Interfaces and surfaces have been the focus of intense research in the past few decades from both points of pure science and for the creation of new devices\cite{hellman2017interface,soumyanarayanan2016emergent,lee2016direct,luo2021recent,bai2015surface,gao2020recent}. This increased interest in interfaces and surfaces is due to the discovery of new properties and new phases non-existing in the bulk and to the progress from a technological point that allows to handle and manipulate materials in low dimensions and low thicknesses\cite{li2021surface,liu2018interface,butt2023physics}. 
% altermagnetism, applications and space groups
Very recently, the spin-splitting in the electronic bands, typical of ferromagnets, was found in commensurate compounds with crystal-symmetry compensated magnetic order.\cite{mazin2022altermagnetism} This magnetic phase was called altermagnetism (AM)\cite{hayami2019momentum,hayami2020bottom,Smejkal22beyond,mazin2022altermagnetism}or antiferromagnetism with non-interconvertible spin-structure motif pair\cite{yuan2023degeneracy}. 

\begin{figure}[ht!]
\centering
\includegraphics[width=0.98\linewidth]{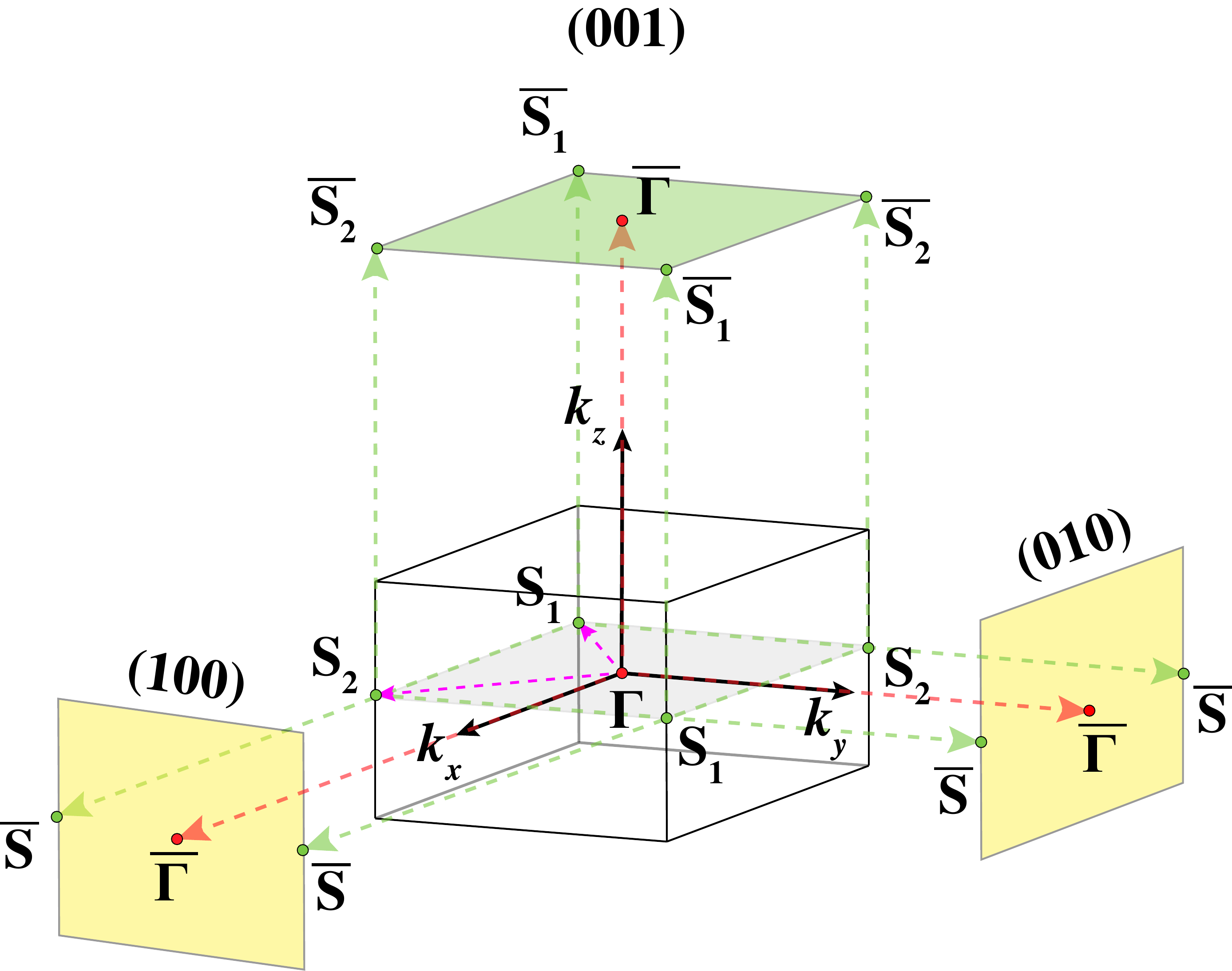}
\caption{Brillouin zone with high-symmetry points for the tetragonal RuO$_2$. With subscripts 1 and 2, we indicate the two points in the $k$-space that have opposite non-relativistic spin-splitting that are S$_1$ and S$_2$ in this example. In magenta, we highlight the $k$-path S$_1$-$\Gamma$-S$_2$ where the altermagnetic spin-splitting is maximized. We project the bulk Brillouin zone on the principal surfaces (100), (010) and (001).  The projected high-symmetry points have an overline. Given the geometrical position of the $k$-points with opposite non-relativistic spin-splitting, the altermagnetic surface states survived on the (001) surface (colored in green), while the other two surfaces (colored in yellow) are blind to AM. The surfaces blind to AM are simple antiferromagnetic surfaces with double degeneracy. The projected high-symmetry points have a subscript if the band structure connecting $\overline{\Gamma}$ still preserves the altermagnetic properties, they do not show subscript otherwise.}\label{BZ-schematic}
\end{figure}

The presence of AM requires the electronic charge of the spin-up (down) atoms to be mapped in the charge of the spin-down (up) atoms 
without translations, inversion, or their combinations, but only with roto-translations or mirrors\cite {Smejkal22,Smejkal22beyond}. 
Therefore, it is observable for some determined space groups and even more precisely, in the magnetic space groups of type-I and type-III\cite{GUO2023100991}. 
If the system is metallic, the AM systems can exhibit spontaneous anomalous Hall effects (AHE) even in the absence of net magnetic moment due to the presence of a non-relativistic spin-splitting in the momentum space\cite{doi:10.1126/sciadv.aaz8809,PhysRevLett.130.036702}.
The direction of the Hall vector is governed by the N\'eel vector defined as the difference between the magnetization vectors on the two different antiferromagnetic sublattices.\cite{turek2022altermagnetism,shao2023neel,Fakhredine23,Sanyal2023}

Regarding technological applications, altermagnets may assume a leading role in realizing concepts in spincaloritronics\cite{zhou2023crystal}, THz emissions in spintronic systems\cite{liuinverse}, THz spin current dynamics in g-wave altermagnet hematite\cite{qiu2023terahertz} and an efficient spin-charge conversion\cite{bai2023efficient}. Furthermore, they can be used in Josephson junctions\cite{Ouassou23} and to generate neutral currents for spintronics\cite{Shao21}. However, there is scope for further exploration to identify techniques to enhance domain structure control in altermagnetic systems as far as spintronic applications are concerned.\cite{rimmler2023atomic} As a solution to control altermagnetic compounds, circularly polarized light can be employed which exploits the \textit{unique} magneto-optical responses for efficient detection, induction and switching applications.\cite{ahn2023flipping}

% what we will do in this paper
In this work, our objective is to have a general understanding of the evolution of the AM properties going toward surfaces and interfaces. Since AM strongly depends on the magnetic space groups, we expect a not straightforward behavior on the surfaces where dimensionality is reduced and the symmetries get affected.
A hint that the dimensionality is relevant for the altermagnetic properties was obtained from the orbital-selective AM in the quasi-two-dimensional Ca$_2$RuO$_4$ system, where the three-dimensional d$_{xz}$/d$_{yz}$ bands show AM spin-splitting while the two-dimensional d$_{xy}$ bands do not\cite{Cuono23orbital}.
More in detail, in this paper we describe the evolution of AM on the surfaces and interfaces projecting the 3D Brillouin zone on the 2D Brillouin zone. A representative example is reproduced in Fig. \ref{BZ-schematic}. Considering the principal surface orientations, the AM survives on one surface and gets annihilated on the other two orientations due to the merging of the $k$-points with opposite altermagnetic spin-splitting. 
Which surface is altermagnetic depends on the magnetic space group, the magnetic order and the details of the crystal structure. 
Here, we calculate the surface states for three of the most common space groups in which AM is present, the orthorhombic space group no. 62, the hexagonal space group no. 194 and the tetragonal space group no. 136. Additionally, the space group no. 62 hosts nonsymmorphic symmetries and it was recently shown that a large AHE can be produced in this space group in the case of metals.\cite{PhysRevB.107.155126,Fakhredine23}
The paper is organized as follows: the second section is dedicated to the results and discussion. Three subsections are dedicated to the results of the three different space groups and a fourth subsection is devoted to the discussion. Finally, the authors draw conclusions with future prospects.

\section{Results and discussion}

\begin{figure}[ht!]
\centering
\includegraphics[width=1\linewidth]{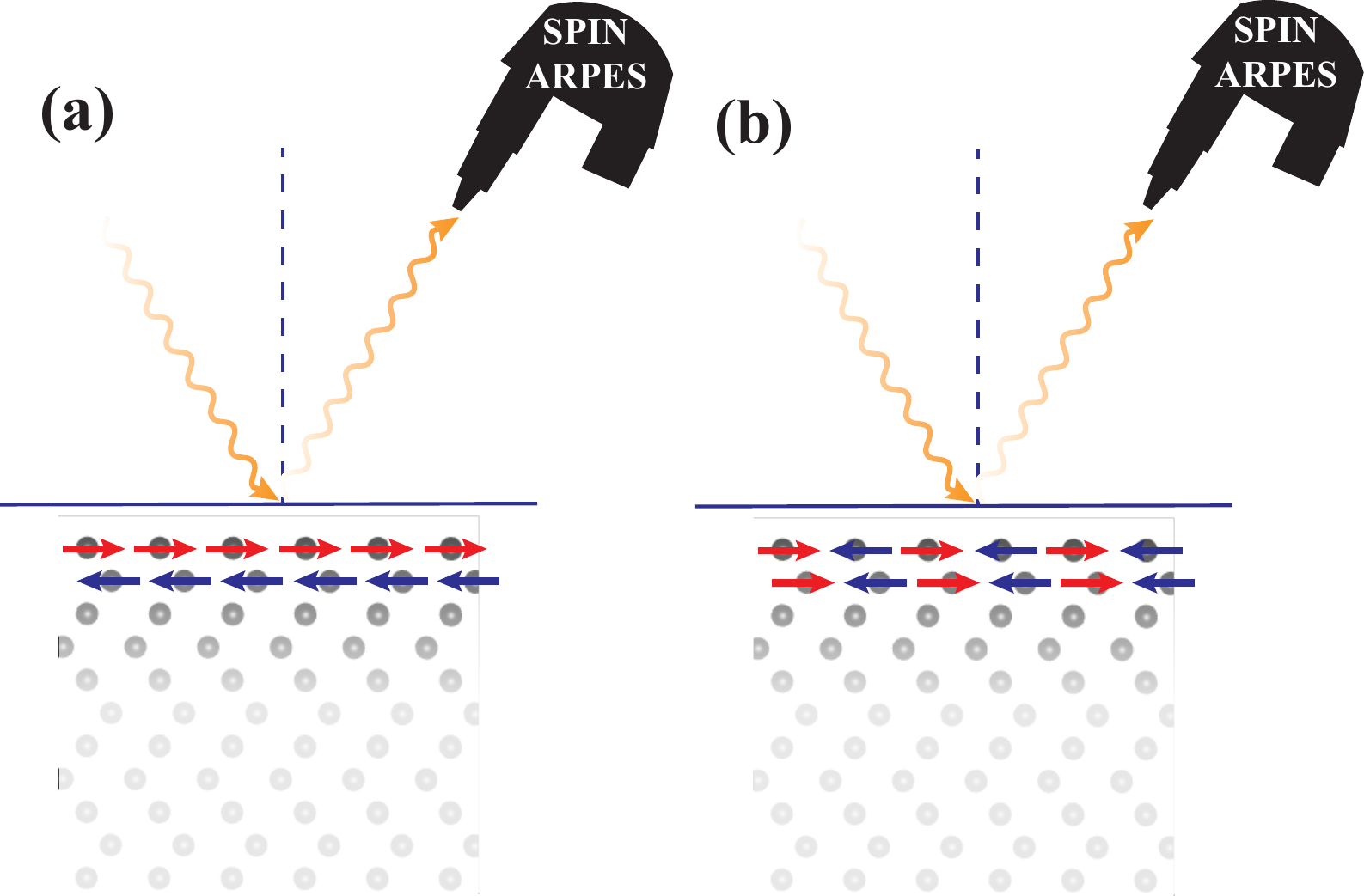}
\caption{(a) Schematic representation of a spin-polarized surface, the surface layer has a net magnetic moment. (b) Schematic representation of a non spin-polarized surface, the surface layer has zero total magnetic moment.}\label{spinpolarized}
\end{figure}

Before studying the altermagnetic surface states, we need to divide the surfaces of the antiferromagnets into two categories depending on the interplay between the magnetic order and the surface. We have the spin-polarized surface as presented in the schematic Fig. \ref{spinpolarized}(a) and the non spin-polarized surface shown in the schematic Fig. \ref{spinpolarized}(b).

The spin-polarized surfaces cannot be investigated for the study of the altermagnetism because the spin-polarization of the surface will be confused with the spin-polarization from the altermagnetism. Therefore, experimental studies should be focused on the non spin-polarized surfaces. We will see that the (001) surfaces of MnTe and A-type LaMnO$_3$ are spin-polarized and therefore are not suitable for experimental investigation of altermagnetism.

\subsection*{Orthorhombic space group Pbnm(62): the case of LaMnO$_3$}

In this first subsection, we investigate the altermagnetic surface states of the space group no. 62.
As a testbed compound for this space group, we choose LaMnO$_3$ which belongs to the large family of perovskites\cite{AUTIERI2023414407}.
The experimental magnetic ground state of the LaMnO$_3$ is the A-type AFM. Therefore, we consider LaMnO$_3$ within this magnetic order and calculate the altermagnetic surface states without relativistic effects (see Supporting Information (SI) section-I for the computational details). 

\begin{figure}[t!]
\centering
\includegraphics[width=1\linewidth]{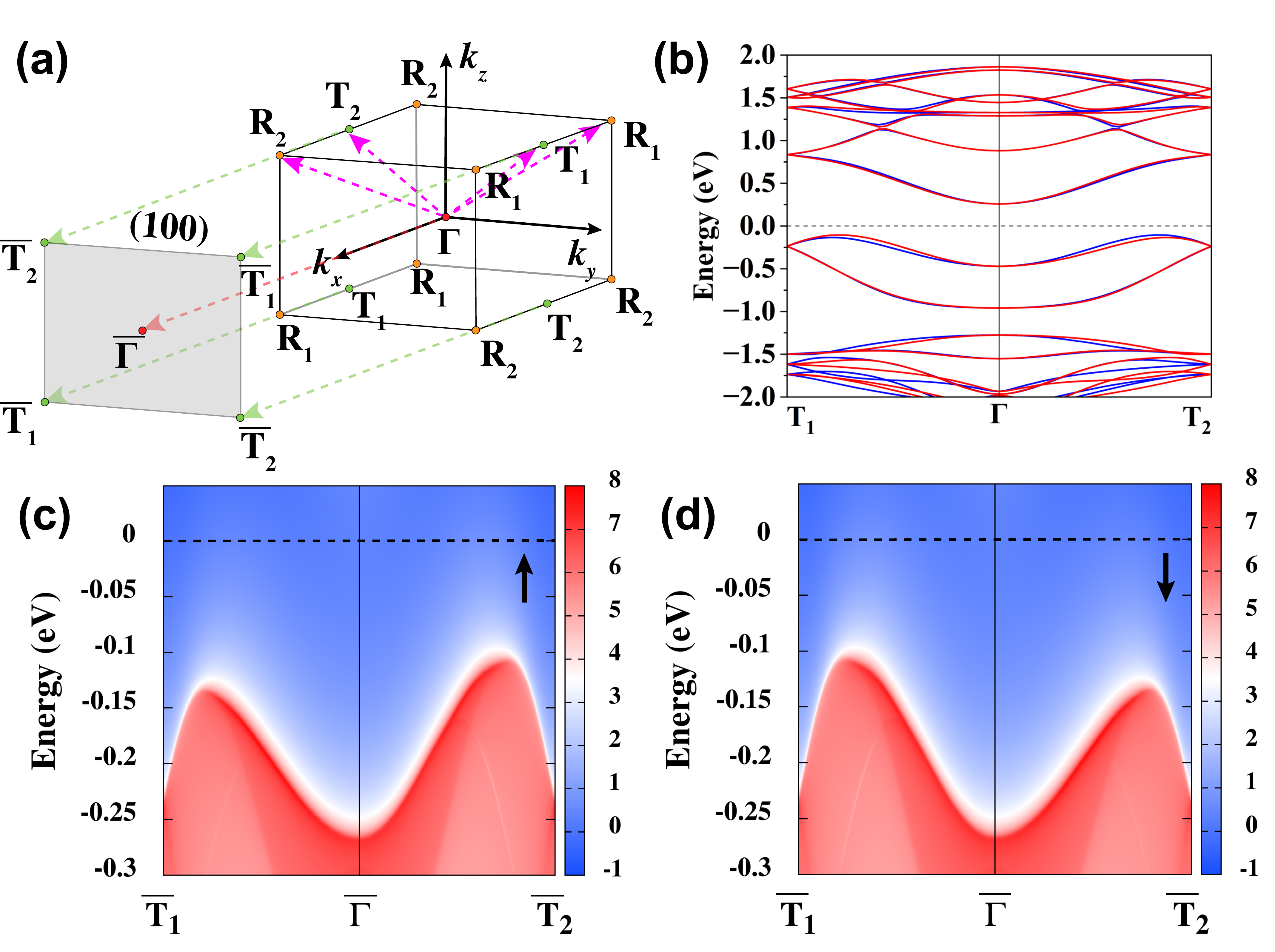}
\caption{(a) Symmetries of the Brillouin zone for the orthorhombic LaMnO$_3$ with the position of R$_1$, R$_2$, T$_1$ and T$_2$ and their projections on the (100) surface orientation. In magenta, we highlight the $k$-paths T$_1$-$\Gamma$-T$_2$ and R$_1$-$\Gamma$-R$_2$ where the altermagnetic spin-splitting is maximized.
(b) Bulk band structure along the $k$-path T$_1$-$\Gamma$-T$_2$, in blue we represent the spin-up channel and in red we represent the spin-down channel. Altermagnetic surface states on the (100) surface orientation for the (c) spin-up and (d) spin-down channel. The Fermi level is set to zero. In the surface band structure, the red color means large spectral weight while the blue color means zero spectral weight.}\label{LaMnO3-ss-bands}
\end{figure}

We plot the symmetries of the Brillouin zone in Fig. \ref{LaMnO3-ss-bands}(a). With subscripts 1 and 2, we indicate the two points in the $k$-space that have opposite non-relativistic spin-splitting, namely the couple of $k$-points R$_1$/R$_2$ and T$_1$/T$_2$ in this example.
The positions of the R$_1$/R$_2$ and T$_1$/T$_2$ points are difficult to predict a priori but they can be easily calculated within first-principle calculations of the band structure from $\Gamma$ towards all the equivalent points. The positions of the R$_1$ and R$_2$ points strongly depend on the space group, the magnetic order\cite{Cuono23orbital} and the details of the crystal structure as the Wyckoff positions of the magnetic atoms\cite{Fakhredine23}.
To study the altermagnetic surface states on the principal surface orientations, we project the positions of R$_1$, R$_2$, T$_1$ and T$_2$ on that surfaces. 
In general, when the $k$-points with opposite non-relativistic spin-splitting merge on the projected surface Brillouin zone, the AM gets annihilated as shown in Fig. \ref{BZ-schematic}. From the projection of the 3D Brillouin zone on the 2D surface Brillouin zone of the A-type phase of LaMnO$_3$, we obtain that the (010) and (001) surfaces are blind to the AM and host double-degenerate antiferromagnetic surface states. 
On the (100) projected Brillouin zone, R$_1$ does not merge with R$_2$. On the contrary, the $k$-paths R$_1$-$\Gamma$-R$_2$ and T$_1$-$\Gamma$-T$_2$ merge in the surface projected $\overline{\rm T}_1$-$\overline{\Gamma}$-$\overline{\rm T}_2$ $k$-path as shown in Fig. \ref{LaMnO3-ss-bands}(a).
Therefore, we study the (100) surface orientation where we expect to find the altermagnetic surface states.
The bulk band structure of LaMnO$_3$ along the path T$_1$-$\Gamma$-T$_2$ is represented in Fig. \ref{LaMnO3-ss-bands}(b), the altermagnetic spin-splitting at half-way between $\Gamma$ and the T points is of the order of 30 meV in the valence band maxima. The band structure of the spin-up channel along $\Gamma$-T$_1$ (or T$_2$) is the same as the band structure of the spin-down channel along $\Gamma$-T$_2$ (or T$_1$).
The altermagnetic surface states for spin-up and spin-down are represented in Fig. \ref{LaMnO3-ss-bands}(c,d), respectively.
The altermagnetic properties are entirely preserved on the (100) surfaces and even the size of the non-relativistic spin-splitting on the surface is unchanged with respect to the bulk spin-splitting.
Indeed, analogously to the bulk, the (100) surface band structure of the spin-up channel along $\overline{\Gamma}$-$\overline{\rm T}_1$ 
(or $\overline{\rm T}_2$) is the same as the band structure of the spin-down channel along $\overline{\Gamma}$-$\overline{\rm T}_2$ 
(or $\overline{\rm T}_1$).
The symmetries of the Brillouin zone depend on the magnetic order, a different magnetic order will change the position of R$_1$ and R$_2$ varying the surfaces that are blind or not to the AM\cite{Cuono23orbital}.

The (001) surface of the A-type magnetic order is spin-polarized.
The Mn atoms with opposite spins are stacked along the z-axis and are connected by a vector (0,0,c/2).
When we create the slab with the tight-binding model, we have the (001) orientation with two different terminations. In this case, the Mn$_\uparrow$ and Mn$_\downarrow$ atoms are equivalent with respect to the surfaces (100) and (010).
More generally, if the vector normal to the surface is orthogonal to the connecting vector between spin-up and spin-down, therefore, spin-up and spin-down are equivalent. 
In the case of the (001) surface, the slab is composed of alternated MnO$_2$ and LaO layers with spins alternating up and down for the A-type magnetic order. The slab will be composed of the following order of layers Mn$_\uparrow$O$_2$/LaO/..../Mn$_\downarrow$O$_2$/LaO. Therefore, in the (001) surface the Mn$_\uparrow$ and Mn$_\downarrow$ are inequivalent since Mn$_\uparrow$ is on the surface and Mn$_\downarrow$ on the subsurface. Increasing the number of layers, this difference gets reduced but it does not vanish leaving the spin-up and spin-down band structure slightly inequivalent but this is completely unrelated to the AM. We define this case as surface uncompensated antiferromagnetism. This surface uncompensated antiferromagnetism is described in SI(section-III).

\subsection*{Hexagonal space group P6$_3$/mmc(194): the case of MnTe}

The structural ground state of bulk MnTe is the $\alpha$-phase with a hexagonal crystal structure \cite{PhysRevB.96.214418,Bossini_2020}
that is magnetic below Néel temperature $T_{\text{N}}= 310$\,K  \cite{PhysRevB.96.214418}.
MnTe is one of the prototype systems for AM \cite{Smejkal22,Smejkal22beyond,Mazin23,Aoyama23,Hariki23} due to the large spin-splitting, large AHE due to the intrinsic p-doping\cite{PhysRevLett.130.036702}, and strongly sensitive valence bands to the orientations of magnetic moments\cite{junior2023sensitivity}.

\begin{figure}[t!]
\centering
\includegraphics[width=1\linewidth]{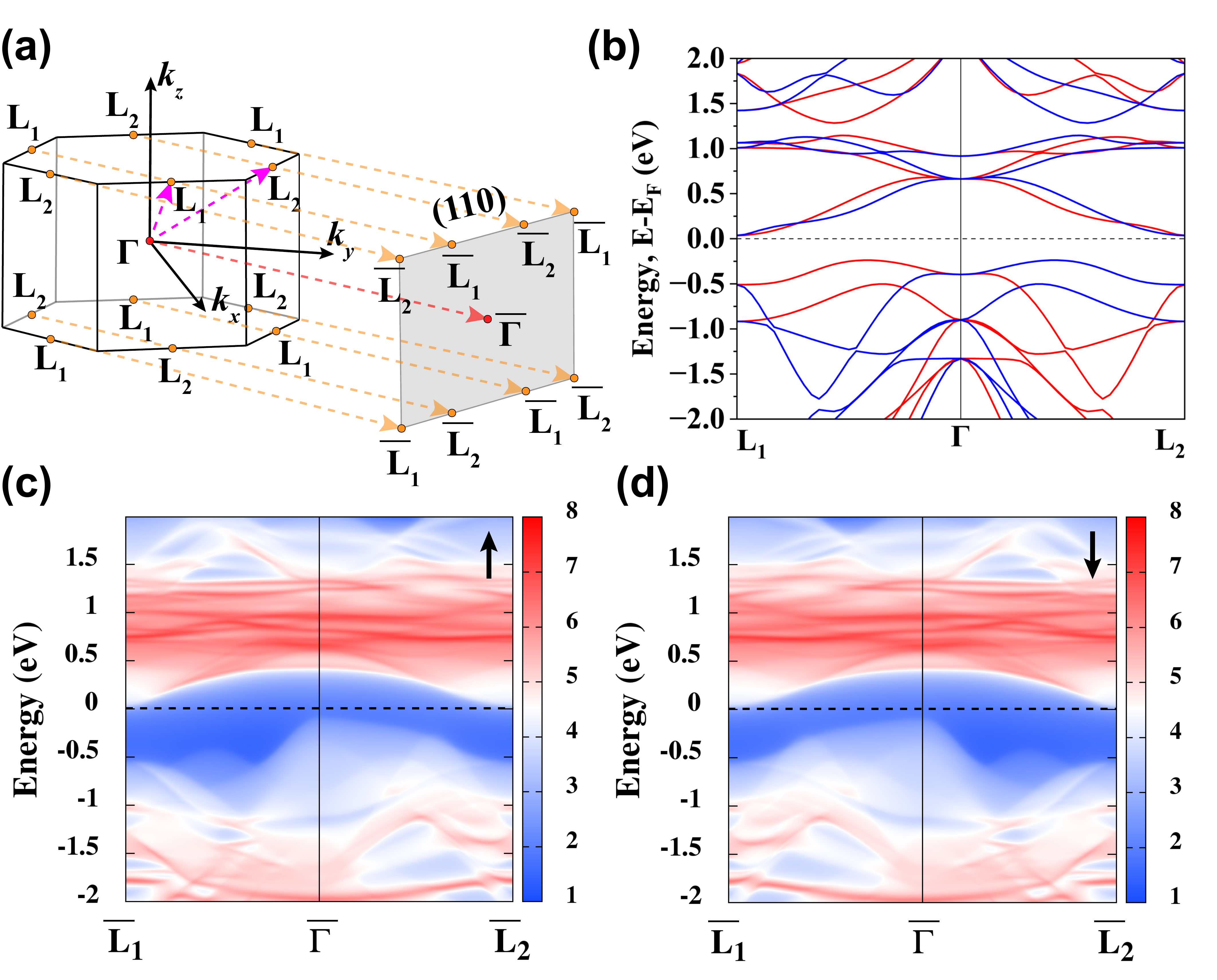}
\caption{(a) Symmetries of the Brillouin zone for hexagonal MnTe with the position of L$_1$ and L$_2$ points and their projections on the (110) surface orientation. In magenta, we highlight the $k$-path L$_1$-$\Gamma$-L$_2$ where the altermagnetic spin-splitting is maximized. (b) Bulk band structure along the $k$-path L$_1$-$\Gamma$-L$_2$, in blue we represent the spin-up channel and in red we represent the spin-down channel. Altermagnetic surface states on the (110) surface orientation for the (c) spin-up and (d) spin-down channel. The Fermi level is set to zero. In the surface band structure, the red color means large spectral weight while the blue color means zero spectral weight.}\label{MnTe-ss-bands}
\end{figure}

The symmetries of the Brillouin zone for hexagonal MnTe are shown in Fig. \ref{MnTe-ss-bands}(a). With subscripts 1 and 2, we indicate the two points in the $k$-space that have opposite non-relativistic spin-splitting.
We project the positions of L$_1$ and L$_2$ points on the principal surface orientations in order to investigate the AM.
At the surfaces (1$\overline{1}$0) and (001), the L$_1$ and L$_2$ points merge, therefore AM is annihilated and these surfaces are blind to altermagnetism.
On the contrary, the (110) surface states present AM because L$_1$ does not merge with L$_2$, as it is shown in Fig. \ref{MnTe-ss-bands}(a).  %The (110) surface shows a zigzag surface in the real space.
Also in this case as in the LaMnO$_3$, the (001) surface is spin-polarized since the magnetic atoms with opposite spin are connected by a vector (0,0,c/2). The (001) surface shows surface uncompensated magnetism in the same fashion as for LaMnO$_3$ (see SI(section-III)).
The bulk band structure of MnTe along the path L$_1$-$\Gamma$-L$_2$, where the altermagnetic spin-splitting is present, is represented in Fig. \ref{MnTe-ss-bands}(b). The band structure of the spin-up channel along $\Gamma$-L$_1$ (or L$_2$) is the same as the band structure of the spin-down channel along $\Gamma$-L$_2$ (or L$_1$).
The same properties of the bulk are preserved in the 
altermagnetic surface states on the (110) surface for spin-up and spin-down channels as shown in Fig. \ref{MnTe-ss-bands}(c,d), respectively. Analogously to the bulk, the (110) surface band structure of the spin-up channel along $\overline{\Gamma}$-$\overline{\rm L}_1$ 
(or $\overline{\rm L}_2$) is the same as the band structure of the spin-down channel along $\overline{\Gamma}$-$\overline{\rm L}_2$ 
(or $\overline{\rm L}_1$).
\\

Several investigations were done on the interface between CrSb or MnTe (belonging to the same space group) and topological insulators in order to create axion insulators.\cite{PhysRevLett.120.056801,lei2021large,PhysRevB.103.195308}
Since the (001) surface of MnTe is blind to AM, the AM is not present in the (001) heterostructures of topological insulators and MnTe that were previously studied\cite{PhysRevB.103.195308}.

\subsection*{Tetragonal space group P4$_2$/mnm(136): the case of RuO$_2$}

\begin{figure}[t!]
\centering
\includegraphics[width=1\linewidth]{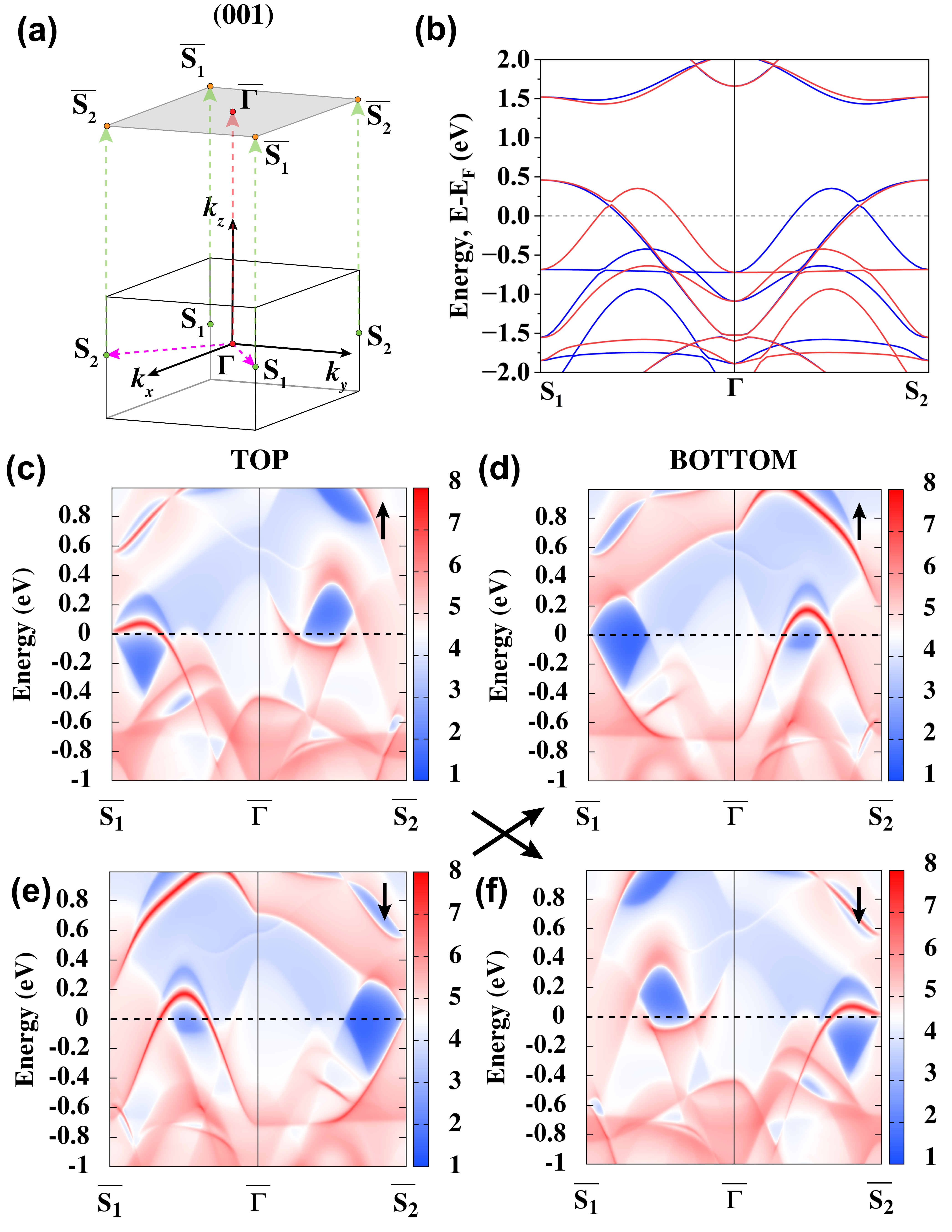}
\caption{(a) Symmetries of the Brillouin zone for the tetragonal RuO$_2$ with the position of S$_1$ and S$_2$ points and their projections on the (001) surface orientation. In magenta, we report the k-path S$_1$-$\Gamma$-S$_2$ where the altermagnetic spin-splitting is maximized.
(b) Bulk band structure along the $k$-path S$_1$-$\Gamma$-S$_2$, in blue we represent the spin-up channel and in red we represent the spin-down channel. Altermagnetic surface states of the (001) surface orientation for the (c,d) spin-up (top and bottom surfaces) and (e,f) spin-down channel (top and bottom surfaces). The AM is sensitive to surface termination as evident here in the top and bottom surface of the slab. The Fermi level is set to zero. In the surface band structure, the red color means large spectral weight while the blue color means zero spectral weight.}\label{RuO2}
\end{figure}

In this subsection, we calculate the altermagnetic surface states of the RuO$_2$ compound which belongs to the space group no. 136 and it is one of the most studied systems in the field of AM  owing to its large spin-splitting, large N\'eel temperature and metallicity leading to the AHE\cite{Smejkal22,Smejkal22beyond}.

Here, we report the Brillouin zone and the electronic properties in Fig. \ref{RuO2}. From the symmetries of the Brillouin zone in Fig. \ref{RuO2}(a), we derive the presence of altermagnetic surface states for the (001) surface orientation. The altermagnetic surface states are absent on the two other principal surface orientations since the
$k$-points with opposite non-relativistic spin-splitting characters annihilate each other when projected on the surfaces making the (100) and (010) surfaces blind to AM. The non-relativistic spin-splitting observed along the $k$-path S$_1$-$\Gamma$-S$_2$ in the electronic bulk properties presented in Fig. \ref{RuO2}(b) is in agreement with the literature. 

\begin{figure}[ht!]
\centering
\includegraphics[width=1\linewidth]{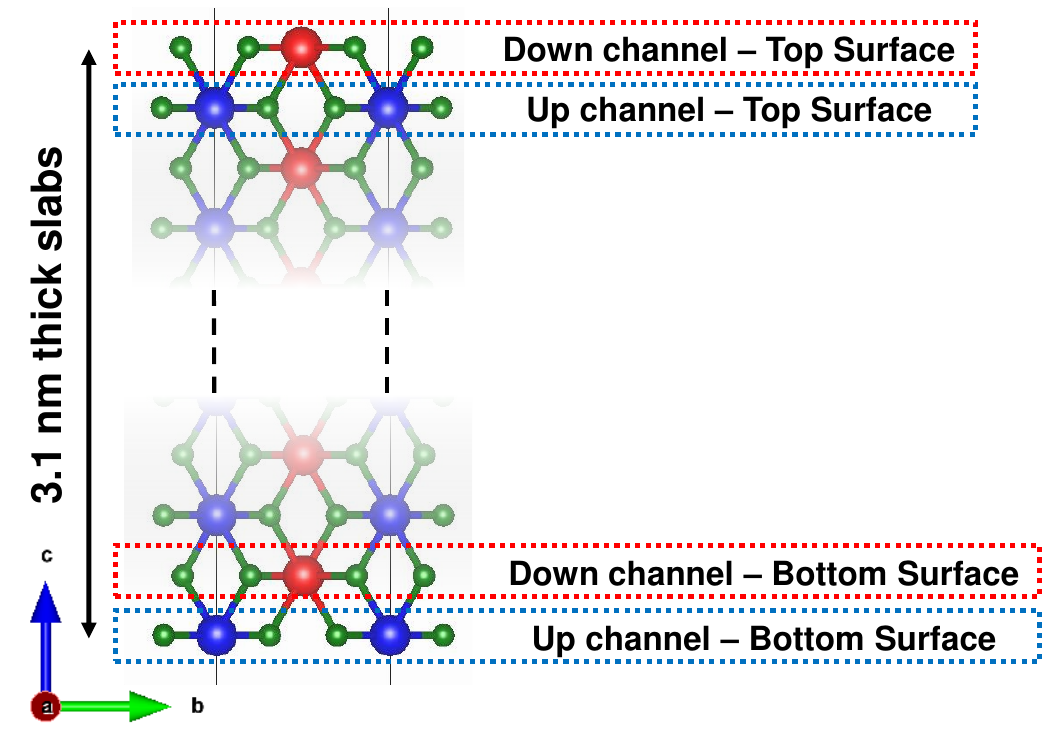}
\caption{Slab structure of RuO$_2$ with (001) surface orientation. We have the equivalent top and bottom surfaces terminating with RuO$_2$ for spin-up and spin-down channels with a 3.11 nm thickness slab. The red and blue balls represent the Ru$_\downarrow$ and Ru$_\uparrow$, respectively. The green balls represent the oxygen atoms.}\label{slab_tetragonal}
\end{figure}

As compared to LaMnO$_3$ and MnTe systems discussed earlier, the altermagnetic surface states of RuO$_2$ have different properties as evident from Fig. \ref{RuO2}(c-f). This nature of altermagnetic surface states originates from the different terminations of the slab presented in Fig. \ref{slab_tetragonal} wherein the top surface is occupied with Ru$_\downarrow$ atoms, while the bottom surface is occupied with the Ru$_\uparrow$ atoms. Whereas the sub-surfaces are opposite with the top subsurface occupied with Ru$_\uparrow$ and the bottom subsurface is Ru$_\downarrow$. As a result, the surface states originating from the Ru$_\downarrow$ atoms on the top surface are altermagnetic partners of the surface states originating from the Ru$_\uparrow$ atoms on the bottom surface. Differently from RuO$_2$, in the LaMnO$_3$ and MnTe, the terminations of the altermagnetic surface contain both atoms with majority spin-up and spin-down.

\subsection*{Electric field control of the surface states}

\begin{figure}[h!]
\centering
\includegraphics[width=1\linewidth]{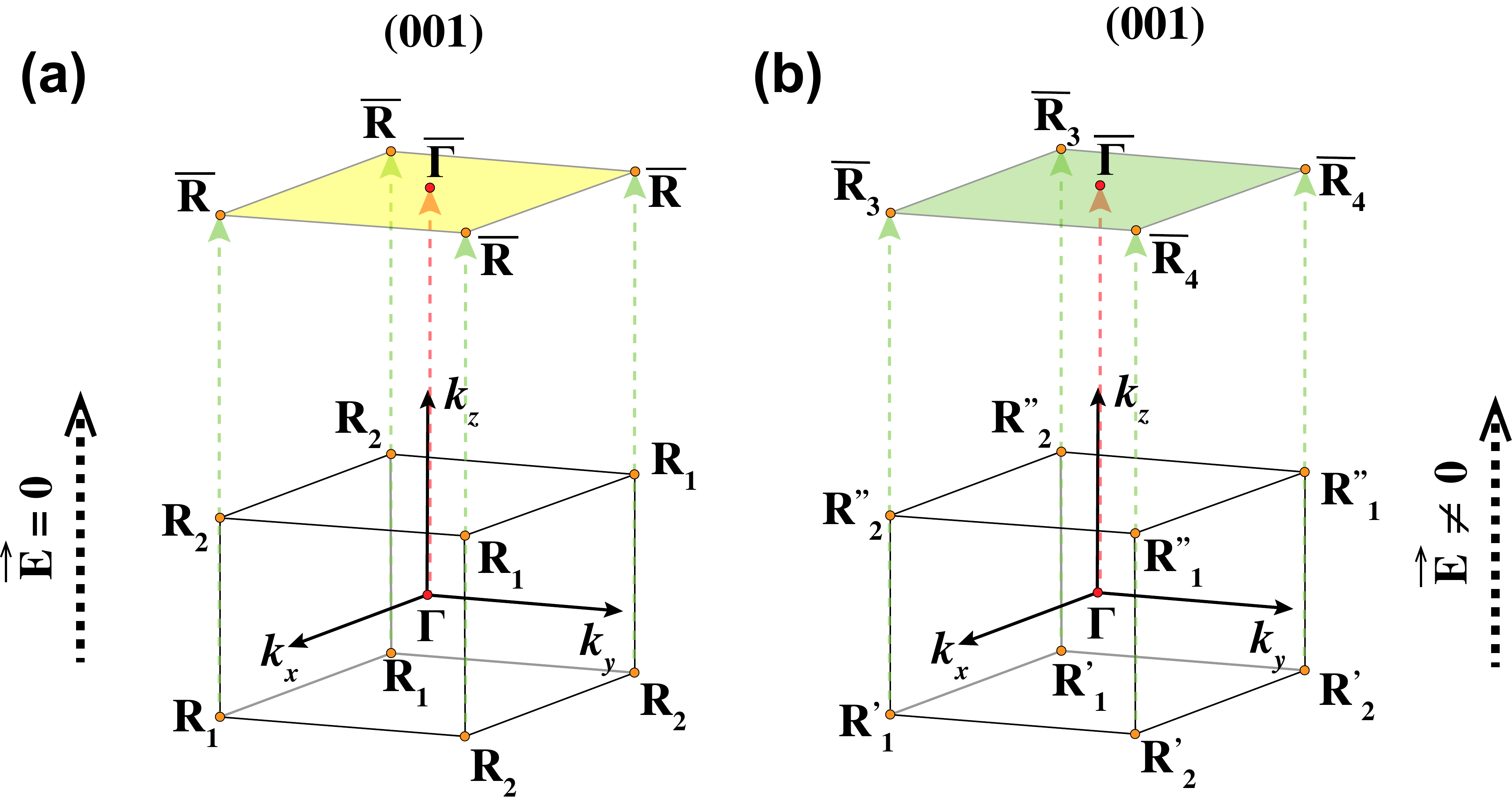}
\caption{(a) In the absence of an external electric field ($\Vec{E} =$ 0) AM is annihilated on the (001) surface since $\overline{R}$ contains the projections of R$_1$ and R$_2$. However, in the presence of an external electric field ($\Vec{E} \neq$ 0) the altermagnetic surface states are activated on the (001) surface. We define R$_3$ as the point that contains the projections of R$_1^{,}$ and R$_2^{,,}$ while R$_4$ contains the projections of R$_2^{,}$ and R$_1^{,,}$.}\label{electric_field}
\end{figure}

Let us consider the surfaces without spin-polarization as described in Fig. \ref{spinpolarized}(b).
We should clarify that the altermagnetic systems do not lose completely all the altermagnetic properties once we create a slab parallel to a blind surface. Indeed, an external electric field perpendicular to the slab can reactivate the altermagnetism. We can understand it from the qualitative behavior of the surface in the presence of an external electric field shown in Fig. \ref{electric_field}. Let us consider the case of the LaMnO$_3$ Brillouin zone, an electric field will break the symmetry related to spin-splitting along $\Gamma$-R$_1$ and $\Gamma$-R$_2$ that are equal and opposite. The electric field along the [001] direction will create R$_1^{,}$ and R$_2^{,}$ points in the bottom of the Brillouin zone while it will create R$_1^{,}$ and R$_2^{,}$ points in the top of the Brillouin zone.
Once we project on the (001) surface, we will have two different points that we define R$_3$ and R$_4$ with different spin-splitting. Therefore, an electric field perpendicular to the blind surface will generate altermagnetism. This will be relevant for the electric control in antiferromagnetic spintronics in the case of insulators. The interplay between electric field and altermagnetism was investigated recently by Guo et al.\cite{guo2023piezoelectric}.

\subsection*{Discussion}

We have proven that there is a strong surface-dependence of the altermagnetic properties. Therefore, every surface or interface study of AM should be performed on a specific surface or interface orientation to observe the effect. Following the recipe given in this work, we can derive the altermagnetic-active surfaces going also beyond the principal surface orientations. For instance, the AM is preserved on the surface (110) of RuO$_2$ as also shown in recent experiments\cite{https://doi.org/10.1002/adom.202300177}.
In the case of alternate positions along all directions of the R$_1$ and R$_2$ points as in the C-type of YVO$_3$ pervoskite\cite{Cuono23orbital}, the altermagnetic surface states will be observable on the (110), (101), (011) and (111) surfaces while all principal surface orientations will be blind to the AM.

As a drawback of this approach, we have to mention that usually on the surfaces and interfaces we have dangling bonds, different magnetic order and/or surface reconstruction that can change the properties of the bulk\cite{PhysRevB.95.115120}. In all our calculations, we assume that there is no surface reconstruction or change of the magnetic order on the surfaces or interface with respect to the bulk. However, if the surface will present just a simple buckling without a change of the magnetic order or the symmetry relevant for the AM, the predictions about the altermagnetic surface states done in this paper are valid.

% prospective
The results presented in this paper are essential for future steps toward the interplay between the AM and topological surface states\cite{Cuono23EuCd2As2}.
Additionally, the schematic representation of symmetries of the Brillouin zone shown in Fig. \ref{BZ-schematic} contains information about the altermagnetic properties for a given space group and magnetic order. 
Further investigations could lead to a faster and easier evaluation of the Hall vector orientation based on the direction of the N\'eel vector and on the symmetries of the Brillouin zone obtained using exclusively DFT results.

\section{Conclusions}
We investigated the surface states of altermagnetic systems considering three representative space groups that host numerous altermagnetic compounds: one tetragonal, one orthorhombic and one hexagonal.
We calculate the 2D projected Brillouin zone from the 3D Brillouin zone and we describe the method to determine the surfaces where the opposite spin-splittings merge annihilating the altermagnetic properties and the surfaces where the AM is preserved. For instance, looking at the three principal surface orientations, we find that two surfaces are blind to AM, while the AM survives for one surface orientation in all considered cases. Where it is preserved, the altermagnetic spin-splitting of the surface states gets unchanged with respect to the bulk. Which surface preserves the AM depends on the relative position of the high-symmetry points (for instance R$_1$ and R$_2$ for the orthorhombic case) in the Brillouin zone. Since the position of these high-symmetry points depends on the magnetic order, also which is the surface hosting the altermagnetic surface states depends on the magnetic order. 
Using an electric field orthogonal to non spin-polarized blind surfaces, we were able to break the inversion symmetry and create altermagnetic surface states on surface orientations that were blind to altermagnetism without field.
   
The results obtained in this paper are a necessary step for further investigations that could lead to a faster and easier evaluation of the Hall vector orientation using exclusively DFT results. Future investigations on the spin texture of the topological surface states are needed in the case of altermagnetic systems.
Our results predict which surfaces to cleave in order to observe altermagnetic spin-splitting in thin films via spin-resolved ARPES. We open future perspectives for the study of altermagnetic effects on the trivial and topological surface states.

\begin{acknowledgments}
We thank T. Dietl and V. V. Volobuev for the useful discussions. The work is supported by the Foundation for Polish Science through the International Research Agendas program co-financed by the European Union within the Smart Growth Operational Programme (Grant No. MAB/2017/1). We acknowledge the access to the computing facilities of the Interdisciplinary Center of Modeling at the University of Warsaw, Grant g91-1418, g91-1419 and g91-1426 for the availability of high-performance computing resources and support. We acknowledge the CINECA award under the ISCRA initiative  IsC99 "SILENTS”, IsC105 "SILENTSG" and IsB26 "SHINY" grants for the availability of high-performance computing resources and support. We acknowledge the access to the computing facilities of the Poznan Supercomputing and Networking Center Grant No. 609.
\end{acknowledgments}

\begin{figure}[h!]
\centering
\includegraphics[width=1\linewidth]{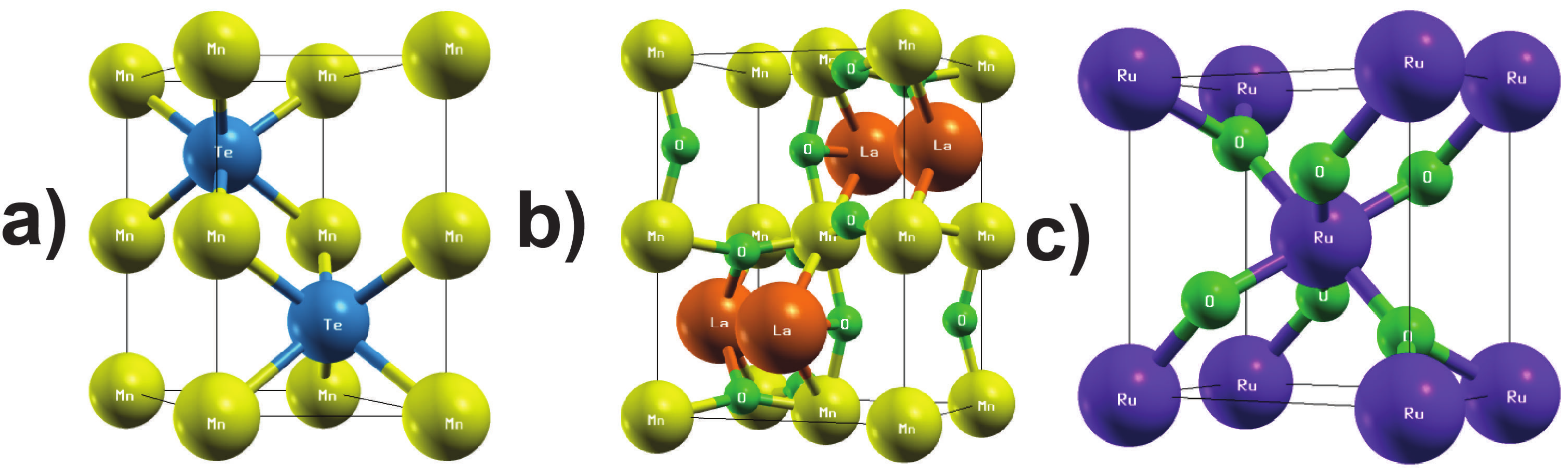}
\caption{Crystal structures for a) hexagonal MnTe (yellow balls are Mn and blue balls are Te), b) orthorhombic LaMnO$_3$ (orange balls are La, yellow balls are Mn, and green balls are O) and c) tetragonal RuO$_2$ (purple balls are Ru and green balls are O) obtained from the materials project repository.\cite{materials-project}}\label{structure}
\end{figure}

\appendix

\section{Computational details}

We performed density functional theory-based \textit{first-principles} calculations as implemented in Quantum ESPRESSO.\cite{giannozzi2009quantum} We performed the calculations without spin-orbit coupling effects. The antiferromagnetic ground state was obtained using ultra-soft pseudopotentials under generalized gradient approximation with Perdew–Burke–Ernzerhof type of exchange-correlation functional.\cite{pseudos,perdew1996generalized} 
%The optimal computational parameters were kept same for all the systems. 
The kinetic energy cut-off of 65 Ry and charge density cut-off of 780 Ry were used with a Monkhorst-Pack grid (\textit{k}-mesh) of 12 $\times$ 12 $\times$ 8.\cite{monkhorst1976special} High electronic self-consistency convergence criteria of at least 10$^{-10}$ were followed in all the calculations. The antiferromagnetic ground state of RuO$_2$ was obtained by implementing Hubbard U with GGA for Ru atoms with \textit{U} $=$ 2 eV and \textit{J$_H$} = 0.15\textit{U}. After these calculations, we performed wannierization for all the systems using Wannier90 code.\cite{w90} The exact tight-binding Hamiltonian generated from wannierization was then used to calculate the altermagnetic surface states using semi-infinite Green's function approach implemented in WannierTools code.\cite{wanniertools}

\begin{figure}[ht!]
\centering
\includegraphics[width=1\linewidth]{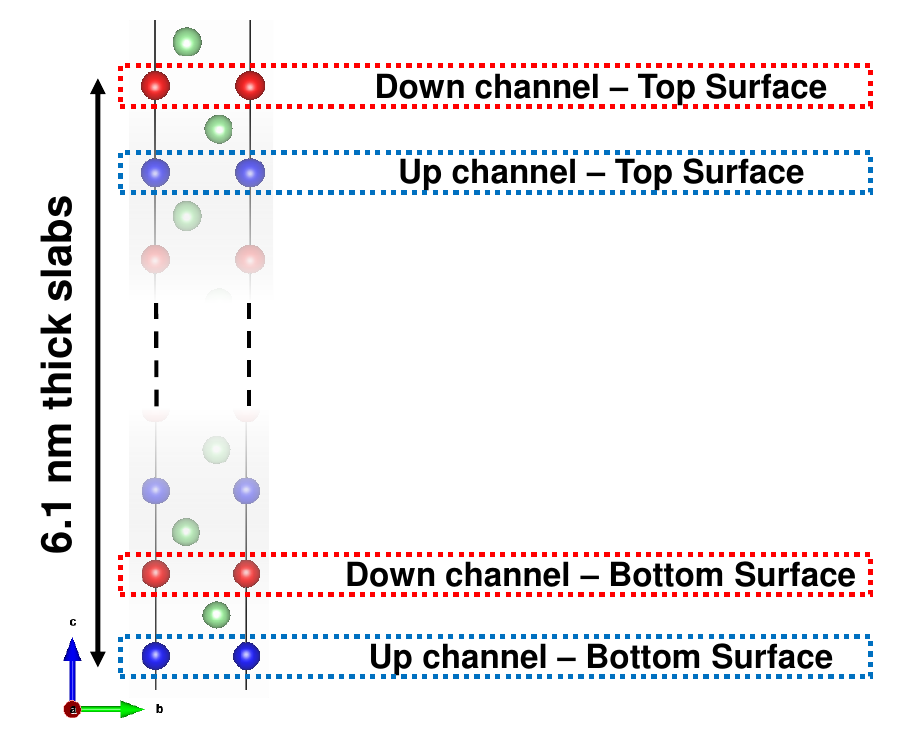}
\caption{Asymmetric and stoichiometric slab of MnTe with (001) surface orientation. Red and blue balls represent the spin-down and spin-up Mn atoms, while green balls represent the Te atoms. The surfaces are inequivalent, the top surface ends with oxygen while the bottom surface ends with Mn.}\label{uncompensated}
\end{figure}

\section{Notations and Structural Details}

Within this paper, (xyz) is the notation to describe the surface plane, while [xyz] is the notation for the direction orthogonal to the surface plane.

We report the crystal symmetries and the structural details of the investigated compounds for a complete understanding of the surface orientations described in the main text. The crystal structures are presented in Fig. \ref{structure}. The Bravais lattice vectors for space group no. 62 are  \textbf{a$_1$}=(a,0,0), \textbf{a$_2$}=(0,b,0) and \textbf{a$_3$}=(0,0,c) while the reciprocal lattice vectors are \textbf{b$_1$}=($\frac{2\pi}{a}$,0,0), \textbf{b$_2$}=(0,$\frac{2\pi}{b}$,0) and \textbf{b$_3$}=(0,0,$\frac{2\pi}{c}$). The Bravais lattice vectors for space group no. 136 are \textbf{a$_1$}=(a,0,0), \textbf{a$_2$}=(0,a,0) and \textbf{a$_3$}=(0,0,c) while the reciprocal lattice vectors are \textbf{b$_1$}=($\frac{2\pi}{a}$,0,0), \textbf{b$_2$}=(0,$\frac{2\pi}{a}$,0) and \textbf{b$_3$}=(0,0,$\frac{2\pi}{c}$). 
The Bravais lattice vectors for space group no. 194 are \textbf{a$_1$}=($\frac{a}{2}$,$\frac{a\sqrt{3}}{2}$,0), \textbf{a$_2$}= ($\frac{a}{2}$,-$\frac{a\sqrt{3}}{2}$,0) and \textbf{a$_3$}= (0,0,c). The reciprocal lattice vectors are \textbf{b$_1$}= ($\frac{2\pi}{a}$,$\frac{2\pi}{a\sqrt{3}}$,0), \textbf{b$_2$}= ($\frac{2\pi}{a}$,$-\frac{2\pi}{a\sqrt{3}}$,0) and \textbf{b$_3$}= (0,0,$\frac{2\pi}{c}$). The [110] $k$-space direction is parallel to the vector \textbf{b$_1$}+\textbf{b$_2$}=($\frac{4\pi}{a}$,0,0).\\

\begin{figure}[h!]
\centering
\includegraphics[width=1\linewidth]{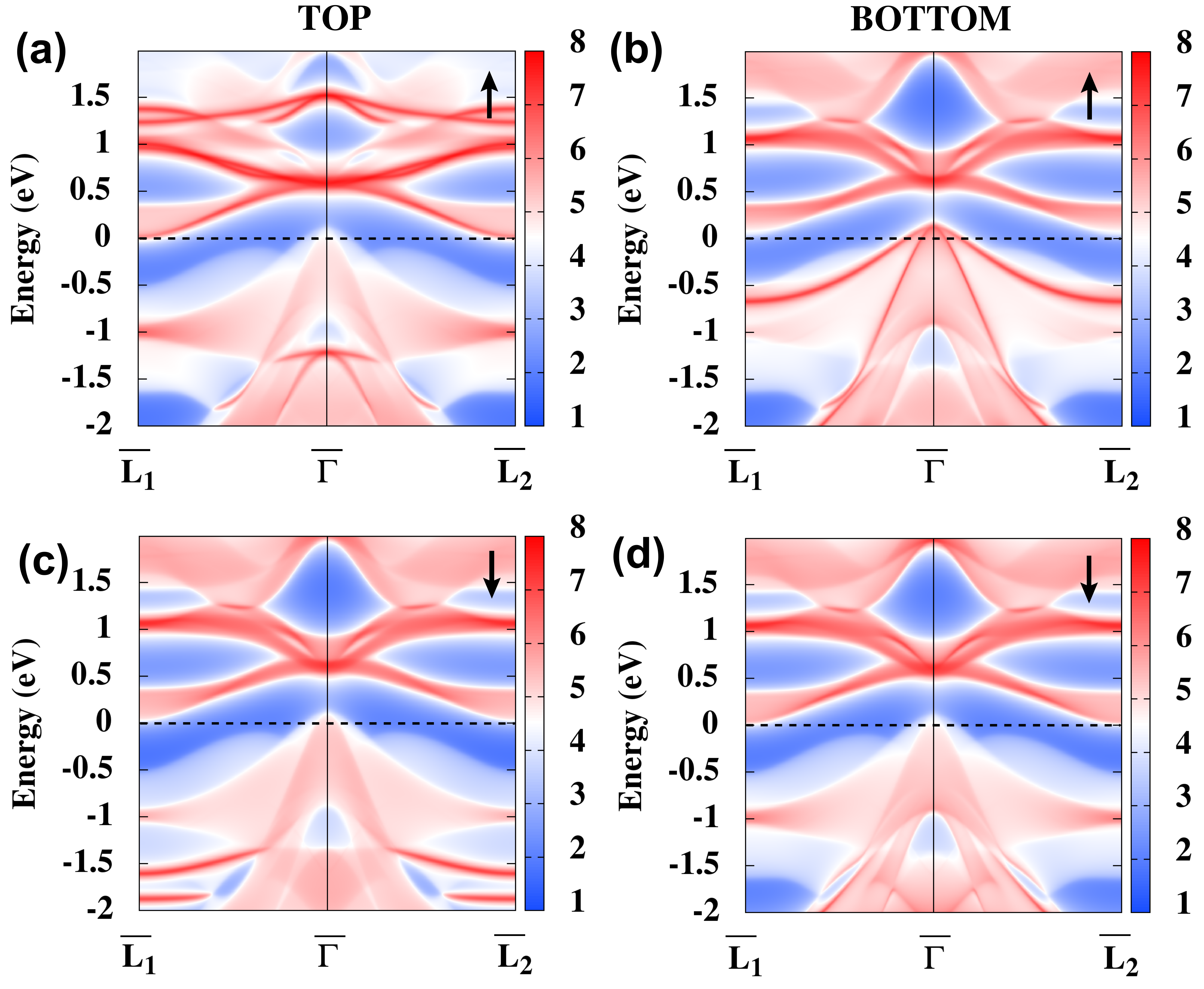}
\caption{Electronic surface states of MnTe along the $k$-path $\overline{\rm L}_1$-$\overline{\Gamma}$-$\overline{\rm L}_2$  for the (a,b) spin-up channel projected on the (001) surface orientation for the (a) top and (b) bottom surfaces, respectively. (c,d) The same for the spin-down channel.
Due to the surface uncompensated magnetism, all these surface band structures are different. The Fermi level is set to zero. In the surface band structure, the red color means large spectral weight while the blue color means zero spectral weight. }\label{MnTe-001}
\end{figure}

The lattice parameters for the three systems under consideration were obtained from the materials project repository \cite{materials-project}.
The optimized lattice parameters used for MnTe were a$=$4.107 {\AA} and c$=$6.467 {\AA}. For LaMnO$_3$ we used the Pbnm setting\cite{AUTIERI2023414407} with a$=$5.585 {\AA}, b$=$5.871 {\AA} and c$=$7.777 {\AA} while for RuO$_2$ we performed the calculations with a$=$b$=$4.482 {\AA} and c$=$3.111 {\AA}. 
MnTe and RuO$_2$ have two magnetic atoms in the unit cells, therefore only one antiferromagnetic configuration is possible. LaMnO$_3$ is an orthorhombic perovskite with the four Mn magnetic atoms in the 4b Wyckoff positions, the A-type magnetic order consists of the 2 Mn atoms at z=0 with spin-up and 2 Mn atoms at the reduced coordinates z=0.5 with spin-down\cite{Cuono23orbital}.

\section{Surface uncompensated magnetism on (001) surface of MnTe}

The spin-up channel and spin-down channel surface states of the (001) surface orientation of MnTe slightly differ due to surface uncompensated magnetism. Indeed, the (001) slab is asymmetric as we can see in Fig. \ref{uncompensated}. The two terminations of the slab (top and bottom) are inequivalent, therefore, the spin-up and spin-down channels are inequivalent.
The surface band structures for the spin-up and spin-down channels and for top and bottom surfaces are reported in Fig. \ref{MnTe-001}. We can observe minor differences among all cases. In the case of DFT simulations with a symmetric slab (that however will not preserve the stoichiometry), we would recover the symmetries observed in the RuO$_2$ case.

The same effect of the surface uncompensated magnetism is present on the (001) surface of LaMnO$_3$, however, the altermagnetic spin-splitting in LaMnO$_3$ is one order of magnitude smaller than MnTe, therefore, this uncompensated magnetism effect is not appreciable in LaMnO$_3$.

\bibliography{altermagnetism}

\begin{thebibliography}{53}%
\makeatletter
\providecommand \@ifxundefined [1]{%
 \@ifx{#1\undefined}
}%
\providecommand \@ifnum [1]{%
 \ifnum #1\expandafter \@firstoftwo
 \else \expandafter \@secondoftwo
 \fi
}%
\providecommand \@ifx [1]{%
 \ifx #1\expandafter \@firstoftwo
 \else \expandafter \@secondoftwo
 \fi
}%
\providecommand \natexlab [1]{#1}%
\providecommand \enquote  [1]{``#1''}%
\providecommand \bibnamefont  [1]{#1}%
\providecommand \bibfnamefont [1]{#1}%
\providecommand \citenamefont [1]{#1}%
\providecommand \href@noop [0]{\@secondoftwo}%
\providecommand \href [0]{\begingroup \@sanitize@url \@href}%
\providecommand \@href[1]{\@@startlink{#1}\@@href}%
\providecommand \@@href[1]{\endgroup#1\@@endlink}%
\providecommand \@sanitize@url [0]{\catcode `\\12\catcode `\$12\catcode
  `\&12\catcode `\#12\catcode `\^12\catcode `\_12\catcode `\%12\relax}%
\providecommand \@@startlink[1]{}%
\providecommand \@@endlink[0]{}%
\providecommand \url  [0]{\begingroup\@sanitize@url \@url }%
\providecommand \@url [1]{\endgroup\@href {#1}{\urlprefix }}%
\providecommand \urlprefix  [0]{URL }%
\providecommand \Eprint [0]{\href }%
\providecommand \doibase [0]{http://dx.doi.org/}%
\providecommand \selectlanguage [0]{\@gobble}%
\providecommand \bibinfo  [0]{\@secondoftwo}%
\providecommand \bibfield  [0]{\@secondoftwo}%
\providecommand \translation [1]{[#1]}%
\providecommand \BibitemOpen [0]{}%
\providecommand \bibitemStop [0]{}%
\providecommand \bibitemNoStop [0]{.\EOS\space}%
\providecommand \EOS [0]{\spacefactor3000\relax}%
\providecommand \BibitemShut  [1]{\csname bibitem#1\endcsname}%
\let\auto@bib@innerbib\@empty
%</preamble>
\bibitem [{\citenamefont {Hellman}\ \emph {et~al.}(2017)\citenamefont
  {Hellman}, \citenamefont {Hoffmann}, \citenamefont {Tserkovnyak},
  \citenamefont {Beach}, \citenamefont {Fullerton}, \citenamefont {Leighton},
  \citenamefont {MacDonald}, \citenamefont {Ralph}, \citenamefont {Arena},
  \citenamefont {D{\"u}rr} \emph {et~al.}}]{hellman2017interface}%
  \BibitemOpen
  \bibfield  {author} {\bibinfo {author} {\bibfnamefont {F.}~\bibnamefont
  {Hellman}}, \bibinfo {author} {\bibfnamefont {A.}~\bibnamefont {Hoffmann}},
  \bibinfo {author} {\bibfnamefont {Y.}~\bibnamefont {Tserkovnyak}}, \bibinfo
  {author} {\bibfnamefont {G.~S.}\ \bibnamefont {Beach}}, \bibinfo {author}
  {\bibfnamefont {E.~E.}\ \bibnamefont {Fullerton}}, \bibinfo {author}
  {\bibfnamefont {C.}~\bibnamefont {Leighton}}, \bibinfo {author}
  {\bibfnamefont {A.~H.}\ \bibnamefont {MacDonald}}, \bibinfo {author}
  {\bibfnamefont {D.~C.}\ \bibnamefont {Ralph}}, \bibinfo {author}
  {\bibfnamefont {D.~A.}\ \bibnamefont {Arena}}, \bibinfo {author}
  {\bibfnamefont {H.~A.}\ \bibnamefont {D{\"u}rr}},  \emph {et~al.},\
  }\href@noop {} {\bibfield  {journal} {\bibinfo  {journal} {Reviews of modern
  physics}\ }\textbf {\bibinfo {volume} {89}},\ \bibinfo {pages} {025006}
  (\bibinfo {year} {2017})}\BibitemShut {NoStop}%
\bibitem [{\citenamefont {Soumyanarayanan}\ \emph {et~al.}(2016)\citenamefont
  {Soumyanarayanan}, \citenamefont {Reyren}, \citenamefont {Fert},\ and\
  \citenamefont {Panagopoulos}}]{soumyanarayanan2016emergent}%
  \BibitemOpen
  \bibfield  {author} {\bibinfo {author} {\bibfnamefont {A.}~\bibnamefont
  {Soumyanarayanan}}, \bibinfo {author} {\bibfnamefont {N.}~\bibnamefont
  {Reyren}}, \bibinfo {author} {\bibfnamefont {A.}~\bibnamefont {Fert}}, \ and\
  \bibinfo {author} {\bibfnamefont {C.}~\bibnamefont {Panagopoulos}},\
  }\href@noop {} {\bibfield  {journal} {\bibinfo  {journal} {Nature}\ }\textbf
  {\bibinfo {volume} {539}},\ \bibinfo {pages} {509} (\bibinfo {year}
  {2016})}\BibitemShut {NoStop}%
\bibitem [{\citenamefont {Lee}\ \emph {et~al.}(2016)\citenamefont {Lee},
  \citenamefont {Katmis}, \citenamefont {Jarillo-Herrero}, \citenamefont
  {Moodera},\ and\ \citenamefont {Gedik}}]{lee2016direct}%
  \BibitemOpen
  \bibfield  {author} {\bibinfo {author} {\bibfnamefont {C.}~\bibnamefont
  {Lee}}, \bibinfo {author} {\bibfnamefont {F.}~\bibnamefont {Katmis}},
  \bibinfo {author} {\bibfnamefont {P.}~\bibnamefont {Jarillo-Herrero}},
  \bibinfo {author} {\bibfnamefont {J.~S.}\ \bibnamefont {Moodera}}, \ and\
  \bibinfo {author} {\bibfnamefont {N.}~\bibnamefont {Gedik}},\ }\href@noop {}
  {\bibfield  {journal} {\bibinfo  {journal} {Nature communications}\ }\textbf
  {\bibinfo {volume} {7}},\ \bibinfo {pages} {12014} (\bibinfo {year}
  {2016})}\BibitemShut {NoStop}%
\bibitem [{\citenamefont {Luo}\ \emph {et~al.}(2021)\citenamefont {Luo},
  \citenamefont {Li}, \citenamefont {Dumont}, \citenamefont {Yu},\ and\
  \citenamefont {Lu}}]{luo2021recent}%
  \BibitemOpen
  \bibfield  {author} {\bibinfo {author} {\bibfnamefont {D.}~\bibnamefont
  {Luo}}, \bibinfo {author} {\bibfnamefont {X.}~\bibnamefont {Li}}, \bibinfo
  {author} {\bibfnamefont {A.}~\bibnamefont {Dumont}}, \bibinfo {author}
  {\bibfnamefont {H.}~\bibnamefont {Yu}}, \ and\ \bibinfo {author}
  {\bibfnamefont {Z.-H.}\ \bibnamefont {Lu}},\ }\href@noop {} {\bibfield
  {journal} {\bibinfo  {journal} {Advanced Materials}\ }\textbf {\bibinfo
  {volume} {33}},\ \bibinfo {pages} {2006004} (\bibinfo {year}
  {2021})}\BibitemShut {NoStop}%
\bibitem [{\citenamefont {Bai}\ \emph {et~al.}(2015)\citenamefont {Bai},
  \citenamefont {Jiang}, \citenamefont {Li},\ and\ \citenamefont
  {Xiong}}]{bai2015surface}%
  \BibitemOpen
  \bibfield  {author} {\bibinfo {author} {\bibfnamefont {S.}~\bibnamefont
  {Bai}}, \bibinfo {author} {\bibfnamefont {W.}~\bibnamefont {Jiang}}, \bibinfo
  {author} {\bibfnamefont {Z.}~\bibnamefont {Li}}, \ and\ \bibinfo {author}
  {\bibfnamefont {Y.}~\bibnamefont {Xiong}},\ }\href@noop {} {\bibfield
  {journal} {\bibinfo  {journal} {ChemNanoMat}\ }\textbf {\bibinfo {volume}
  {1}},\ \bibinfo {pages} {223} (\bibinfo {year} {2015})}\BibitemShut {NoStop}%
\bibitem [{\citenamefont {Gao}\ \emph {et~al.}(2020)\citenamefont {Gao},
  \citenamefont {Zhao}, \citenamefont {Zhang},\ and\ \citenamefont
  {You}}]{gao2020recent}%
  \BibitemOpen
  \bibfield  {author} {\bibinfo {author} {\bibfnamefont {F.}~\bibnamefont
  {Gao}}, \bibinfo {author} {\bibfnamefont {Y.}~\bibnamefont {Zhao}}, \bibinfo
  {author} {\bibfnamefont {X.}~\bibnamefont {Zhang}}, \ and\ \bibinfo {author}
  {\bibfnamefont {J.}~\bibnamefont {You}},\ }\href@noop {} {\bibfield
  {journal} {\bibinfo  {journal} {Advanced Energy Materials}\ }\textbf
  {\bibinfo {volume} {10}},\ \bibinfo {pages} {1902650} (\bibinfo {year}
  {2020})}\BibitemShut {NoStop}%
\bibitem [{\citenamefont {Li}\ \emph {et~al.}(2021)\citenamefont {Li},
  \citenamefont {Liu}, \citenamefont {Dong}, \citenamefont {Gui}, \citenamefont
  {Hu}, \citenamefont {Li},\ and\ \citenamefont {Liu}}]{li2021surface}%
  \BibitemOpen
  \bibfield  {author} {\bibinfo {author} {\bibfnamefont {L.}~\bibnamefont
  {Li}}, \bibinfo {author} {\bibfnamefont {W.}~\bibnamefont {Liu}}, \bibinfo
  {author} {\bibfnamefont {H.}~\bibnamefont {Dong}}, \bibinfo {author}
  {\bibfnamefont {Q.}~\bibnamefont {Gui}}, \bibinfo {author} {\bibfnamefont
  {Z.}~\bibnamefont {Hu}}, \bibinfo {author} {\bibfnamefont {Y.}~\bibnamefont
  {Li}}, \ and\ \bibinfo {author} {\bibfnamefont {J.}~\bibnamefont {Liu}},\
  }\href@noop {} {\bibfield  {journal} {\bibinfo  {journal} {Advanced
  Materials}\ }\textbf {\bibinfo {volume} {33}},\ \bibinfo {pages} {2004959}
  (\bibinfo {year} {2021})}\BibitemShut {NoStop}%
\bibitem [{\citenamefont {Liu}\ and\ \citenamefont
  {Hersam}(2018)}]{liu2018interface}%
  \BibitemOpen
  \bibfield  {author} {\bibinfo {author} {\bibfnamefont {X.}~\bibnamefont
  {Liu}}\ and\ \bibinfo {author} {\bibfnamefont {M.~C.}\ \bibnamefont
  {Hersam}},\ }\href@noop {} {\bibfield  {journal} {\bibinfo  {journal}
  {Advanced Materials}\ }\textbf {\bibinfo {volume} {30}},\ \bibinfo {pages}
  {1801586} (\bibinfo {year} {2018})}\BibitemShut {NoStop}%
\bibitem [{\citenamefont {Butt}\ \emph {et~al.}(2023)\citenamefont {Butt},
  \citenamefont {Graf},\ and\ \citenamefont {Kappl}}]{butt2023physics}%
  \BibitemOpen
  \bibfield  {author} {\bibinfo {author} {\bibfnamefont {H.-J.}\ \bibnamefont
  {Butt}}, \bibinfo {author} {\bibfnamefont {K.}~\bibnamefont {Graf}}, \ and\
  \bibinfo {author} {\bibfnamefont {M.}~\bibnamefont {Kappl}},\ }\href@noop {}
  {\emph {\bibinfo {title} {Physics and chemistry of interfaces}}}\ (\bibinfo
  {publisher} {John Wiley \& Sons},\ \bibinfo {year} {2023})\BibitemShut
  {NoStop}%
\bibitem [{\citenamefont {Mazin}\ \emph {et~al.}(2022)\citenamefont {Mazin}
  \emph {et~al.}}]{mazin2022altermagnetism}%
  \BibitemOpen
  \bibfield  {author} {\bibinfo {author} {\bibfnamefont {I.}~\bibnamefont
  {Mazin}} \emph {et~al.},\ }\href@noop {} {\bibfield  {journal} {\bibinfo
  {journal} {Physical Review X}\ }\textbf {\bibinfo {volume} {12}},\ \bibinfo
  {pages} {040002} (\bibinfo {year} {2022})}\BibitemShut {NoStop}%
\bibitem [{\citenamefont {Hayami}\ \emph {et~al.}(2019)\citenamefont {Hayami},
  \citenamefont {Yanagi},\ and\ \citenamefont {Kusunose}}]{hayami2019momentum}%
  \BibitemOpen
  \bibfield  {author} {\bibinfo {author} {\bibfnamefont {S.}~\bibnamefont
  {Hayami}}, \bibinfo {author} {\bibfnamefont {Y.}~\bibnamefont {Yanagi}}, \
  and\ \bibinfo {author} {\bibfnamefont {H.}~\bibnamefont {Kusunose}},\
  }\href@noop {} {\bibfield  {journal} {\bibinfo  {journal} {journal of the
  physical society of japan}\ }\textbf {\bibinfo {volume} {88}},\ \bibinfo
  {pages} {123702} (\bibinfo {year} {2019})}\BibitemShut {NoStop}%
\bibitem [{\citenamefont {Hayami}\ \emph {et~al.}(2020)\citenamefont {Hayami},
  \citenamefont {Yanagi},\ and\ \citenamefont {Kusunose}}]{hayami2020bottom}%
  \BibitemOpen
  \bibfield  {author} {\bibinfo {author} {\bibfnamefont {S.}~\bibnamefont
  {Hayami}}, \bibinfo {author} {\bibfnamefont {Y.}~\bibnamefont {Yanagi}}, \
  and\ \bibinfo {author} {\bibfnamefont {H.}~\bibnamefont {Kusunose}},\
  }\href@noop {} {\bibfield  {journal} {\bibinfo  {journal} {Physical Review
  B}\ }\textbf {\bibinfo {volume} {102}},\ \bibinfo {pages} {144441} (\bibinfo
  {year} {2020})}\BibitemShut {NoStop}%
\bibitem [{\citenamefont {\ifmmode~\check{S}\else \v{S}\fi{}mejkal}\ \emph
  {et~al.}(2022{\natexlab{a}})\citenamefont {\ifmmode~\check{S}\else
  \v{S}\fi{}mejkal}, \citenamefont {Sinova},\ and\ \citenamefont
  {Jungwirth}}]{Smejkal22beyond}%
  \BibitemOpen
  \bibfield  {author} {\bibinfo {author} {\bibfnamefont {L.}~\bibnamefont
  {\ifmmode~\check{S}\else \v{S}\fi{}mejkal}}, \bibinfo {author} {\bibfnamefont
  {J.}~\bibnamefont {Sinova}}, \ and\ \bibinfo {author} {\bibfnamefont
  {T.}~\bibnamefont {Jungwirth}},\ }\href {\doibase 10.1103/PhysRevX.12.031042}
  {\bibfield  {journal} {\bibinfo  {journal} {Phys. Rev. X}\ }\textbf {\bibinfo
  {volume} {12}},\ \bibinfo {pages} {031042} (\bibinfo {year}
  {2022}{\natexlab{a}})}\BibitemShut {NoStop}%
\bibitem [{\citenamefont {Yuan}\ and\ \citenamefont
  {Zunger}(2023)}]{yuan2023degeneracy}%
  \BibitemOpen
  \bibfield  {author} {\bibinfo {author} {\bibfnamefont {L.-D.}\ \bibnamefont
  {Yuan}}\ and\ \bibinfo {author} {\bibfnamefont {A.}~\bibnamefont {Zunger}},\
  }\href@noop {} {\bibfield  {journal} {\bibinfo  {journal} {Advanced
  Materials}\ ,\ \bibinfo {pages} {2211966}} (\bibinfo {year}
  {2023})}\BibitemShut {NoStop}%
\bibitem [{\citenamefont {\ifmmode~\check{S}\else \v{S}\fi{}mejkal}\ \emph
  {et~al.}(2022{\natexlab{b}})\citenamefont {\ifmmode~\check{S}\else
  \v{S}\fi{}mejkal}, \citenamefont {Sinova},\ and\ \citenamefont
  {Jungwirth}}]{Smejkal22}%
  \BibitemOpen
  \bibfield  {author} {\bibinfo {author} {\bibfnamefont {L.}~\bibnamefont
  {\ifmmode~\check{S}\else \v{S}\fi{}mejkal}}, \bibinfo {author} {\bibfnamefont
  {J.}~\bibnamefont {Sinova}}, \ and\ \bibinfo {author} {\bibfnamefont
  {T.}~\bibnamefont {Jungwirth}},\ }\href {\doibase 10.1103/PhysRevX.12.040501}
  {\bibfield  {journal} {\bibinfo  {journal} {Phys. Rev. X}\ }\textbf {\bibinfo
  {volume} {12}},\ \bibinfo {pages} {040501} (\bibinfo {year}
  {2022}{\natexlab{b}})}\BibitemShut {NoStop}%
\bibitem [{\citenamefont {Guo}\ \emph {et~al.}(2023{\natexlab{a}})\citenamefont
  {Guo}, \citenamefont {Liu}, \citenamefont {Janson}, \citenamefont {Fulga},
  \citenamefont {{van den Brink}},\ and\ \citenamefont
  {Facio}}]{GUO2023100991}%
  \BibitemOpen
  \bibfield  {author} {\bibinfo {author} {\bibfnamefont {Y.}~\bibnamefont
  {Guo}}, \bibinfo {author} {\bibfnamefont {H.}~\bibnamefont {Liu}}, \bibinfo
  {author} {\bibfnamefont {O.}~\bibnamefont {Janson}}, \bibinfo {author}
  {\bibfnamefont {I.~C.}\ \bibnamefont {Fulga}}, \bibinfo {author}
  {\bibfnamefont {J.}~\bibnamefont {{van den Brink}}}, \ and\ \bibinfo {author}
  {\bibfnamefont {J.~I.}\ \bibnamefont {Facio}},\ }\href {\doibase
  https://doi.org/10.1016/j.mtphys.2023.100991} {\bibfield  {journal} {\bibinfo
   {journal} {Materials Today Physics}\ }\textbf {\bibinfo {volume} {32}},\
  \bibinfo {pages} {100991} (\bibinfo {year} {2023}{\natexlab{a}})}\BibitemShut
  {NoStop}%
\bibitem [{\citenamefont {Šmejkal}\ \emph {et~al.}(2020)\citenamefont
  {Šmejkal}, \citenamefont {González-Hernández}, \citenamefont {Jungwirth},\
  and\ \citenamefont {Sinova}}]{doi:10.1126/sciadv.aaz8809}%
  \BibitemOpen
  \bibfield  {author} {\bibinfo {author} {\bibfnamefont {L.}~\bibnamefont
  {Šmejkal}}, \bibinfo {author} {\bibfnamefont {R.}~\bibnamefont
  {González-Hernández}}, \bibinfo {author} {\bibfnamefont {T.}~\bibnamefont
  {Jungwirth}}, \ and\ \bibinfo {author} {\bibfnamefont {J.}~\bibnamefont
  {Sinova}},\ }\href {\doibase 10.1126/sciadv.aaz8809} {\bibfield  {journal}
  {\bibinfo  {journal} {Science Advances}\ }\textbf {\bibinfo {volume} {6}},\
  \bibinfo {pages} {eaaz8809} (\bibinfo {year} {2020})},\ \Eprint
  {http://arxiv.org/abs/https://www.science.org/doi/pdf/10.1126/sciadv.aaz8809}
  {https://www.science.org/doi/pdf/10.1126/sciadv.aaz8809} \BibitemShut
  {NoStop}%
\bibitem [{\citenamefont {Gonzalez~Betancourt}\ \emph
  {et~al.}(2023)\citenamefont {Gonzalez~Betancourt}, \citenamefont
  {Zub\'a\ifmmode~\check{c}\else \v{c}\fi{}}, \citenamefont
  {Gonzalez-Hernandez}, \citenamefont {Geishendorf}, \citenamefont {\ifmmode
  \check{S}\else \v{S}\fi{}ob\'a\ifmmode~\check{n}\else \v{n}\fi{}},
  \citenamefont {Springholz}, \citenamefont {Olejn\'{\i}k}, \citenamefont
  {\ifmmode~\check{S}\else \v{S}\fi{}mejkal}, \citenamefont {Sinova},
  \citenamefont {Jungwirth}, \citenamefont {Goennenwein}, \citenamefont
  {Thomas}, \citenamefont {Reichlov\'a}, \citenamefont {\ifmmode~\check{Z}\else
  \v{Z}\fi{}elezn\'y},\ and\ \citenamefont
  {Kriegner}}]{PhysRevLett.130.036702}%
  \BibitemOpen
  \bibfield  {author} {\bibinfo {author} {\bibfnamefont {R.~D.}\ \bibnamefont
  {Gonzalez~Betancourt}}, \bibinfo {author} {\bibfnamefont {J.}~\bibnamefont
  {Zub\'a\ifmmode~\check{c}\else \v{c}\fi{}}}, \bibinfo {author} {\bibfnamefont
  {R.}~\bibnamefont {Gonzalez-Hernandez}}, \bibinfo {author} {\bibfnamefont
  {K.}~\bibnamefont {Geishendorf}}, \bibinfo {author} {\bibfnamefont
  {Z.}~\bibnamefont {\ifmmode \check{S}\else
  \v{S}\fi{}ob\'a\ifmmode~\check{n}\else \v{n}\fi{}}}, \bibinfo {author}
  {\bibfnamefont {G.}~\bibnamefont {Springholz}}, \bibinfo {author}
  {\bibfnamefont {K.}~\bibnamefont {Olejn\'{\i}k}}, \bibinfo {author}
  {\bibfnamefont {L.}~\bibnamefont {\ifmmode~\check{S}\else \v{S}\fi{}mejkal}},
  \bibinfo {author} {\bibfnamefont {J.}~\bibnamefont {Sinova}}, \bibinfo
  {author} {\bibfnamefont {T.}~\bibnamefont {Jungwirth}}, \bibinfo {author}
  {\bibfnamefont {S.~T.~B.}\ \bibnamefont {Goennenwein}}, \bibinfo {author}
  {\bibfnamefont {A.}~\bibnamefont {Thomas}}, \bibinfo {author} {\bibfnamefont
  {H.}~\bibnamefont {Reichlov\'a}}, \bibinfo {author} {\bibfnamefont
  {J.}~\bibnamefont {\ifmmode~\check{Z}\else \v{Z}\fi{}elezn\'y}}, \ and\
  \bibinfo {author} {\bibfnamefont {D.}~\bibnamefont {Kriegner}},\ }\href
  {\doibase 10.1103/PhysRevLett.130.036702} {\bibfield  {journal} {\bibinfo
  {journal} {Phys. Rev. Lett.}\ }\textbf {\bibinfo {volume} {130}},\ \bibinfo
  {pages} {036702} (\bibinfo {year} {2023})}\BibitemShut {NoStop}%
\bibitem [{\citenamefont {Turek}(2022)}]{turek2022altermagnetism}%
  \BibitemOpen
  \bibfield  {author} {\bibinfo {author} {\bibfnamefont {I.}~\bibnamefont
  {Turek}},\ }\href@noop {} {\bibfield  {journal} {\bibinfo  {journal}
  {Physical Review B}\ }\textbf {\bibinfo {volume} {106}},\ \bibinfo {pages}
  {094432} (\bibinfo {year} {2022})}\BibitemShut {NoStop}%
\bibitem [{\citenamefont {Shao}\ \emph {et~al.}(2023)\citenamefont {Shao},
  \citenamefont {Jiang}, \citenamefont {Ding}, \citenamefont {Zhang},
  \citenamefont {Wang}, \citenamefont {Xiao}, \citenamefont {Gurung},
  \citenamefont {Lu}, \citenamefont {Sun},\ and\ \citenamefont
  {Tsymbal}}]{shao2023neel}%
  \BibitemOpen
  \bibfield  {author} {\bibinfo {author} {\bibfnamefont {D.-F.}\ \bibnamefont
  {Shao}}, \bibinfo {author} {\bibfnamefont {Y.-Y.}\ \bibnamefont {Jiang}},
  \bibinfo {author} {\bibfnamefont {J.}~\bibnamefont {Ding}}, \bibinfo {author}
  {\bibfnamefont {S.-H.}\ \bibnamefont {Zhang}}, \bibinfo {author}
  {\bibfnamefont {Z.-A.}\ \bibnamefont {Wang}}, \bibinfo {author}
  {\bibfnamefont {R.-C.}\ \bibnamefont {Xiao}}, \bibinfo {author}
  {\bibfnamefont {G.}~\bibnamefont {Gurung}}, \bibinfo {author} {\bibfnamefont
  {W.}~\bibnamefont {Lu}}, \bibinfo {author} {\bibfnamefont {Y.}~\bibnamefont
  {Sun}}, \ and\ \bibinfo {author} {\bibfnamefont {E.~Y.}\ \bibnamefont
  {Tsymbal}},\ }\href@noop {} {\bibfield  {journal} {\bibinfo  {journal}
  {Physical Review Letters}\ }\textbf {\bibinfo {volume} {130}},\ \bibinfo
  {pages} {216702} (\bibinfo {year} {2023})}\BibitemShut {NoStop}%
\bibitem [{\citenamefont {Fakhredine}\ \emph {et~al.}(2023)\citenamefont
  {Fakhredine}, \citenamefont {Sattigeri}, \citenamefont {Cuono},\ and\
  \citenamefont {Autieri}}]{Fakhredine23}%
  \BibitemOpen
  \bibfield  {author} {\bibinfo {author} {\bibfnamefont {A.}~\bibnamefont
  {Fakhredine}}, \bibinfo {author} {\bibfnamefont {R.~M.}\ \bibnamefont
  {Sattigeri}}, \bibinfo {author} {\bibfnamefont {G.}~\bibnamefont {Cuono}}, \
  and\ \bibinfo {author} {\bibfnamefont {C.}~\bibnamefont {Autieri}},\
  }\href@noop {} {\bibfield  {journal} {\bibinfo  {journal} {In manuscript}\ }
  (\bibinfo {year} {2023})}\BibitemShut {NoStop}%
\bibitem [{\citenamefont {Chen}\ \emph {et~al.}(2023)\citenamefont {Chen},
  \citenamefont {Wang}, \citenamefont {Li},\ and\ \citenamefont
  {Sanyal}}]{Sanyal2023}%
  \BibitemOpen
  \bibfield  {author} {\bibinfo {author} {\bibfnamefont {X.}~\bibnamefont
  {Chen}}, \bibinfo {author} {\bibfnamefont {D.}~\bibnamefont {Wang}}, \bibinfo
  {author} {\bibfnamefont {L.}~\bibnamefont {Li}}, \ and\ \bibinfo {author}
  {\bibfnamefont {B.}~\bibnamefont {Sanyal}},\ }\href {\doibase
  10.1063/5.0147450} {\bibfield  {journal} {\bibinfo  {journal} {Applied
  Physics Letters}\ }\textbf {\bibinfo {volume} {123}},\ \bibinfo {pages}
  {022402} (\bibinfo {year} {2023})},\ \Eprint
  {http://arxiv.org/abs/https://pubs.aip.org/aip/apl/article-pdf/doi/10.1063/5.0147450/18038936/022402\_1\_5.0147450.pdf}
  {https://pubs.aip.org/aip/apl/article-pdf/doi/10.1063/5.0147450/18038936/022402\_1\_5.0147450.pdf}
  \BibitemShut {NoStop}%
\bibitem [{\citenamefont {Zhou}\ \emph {et~al.}(2023)\citenamefont {Zhou},
  \citenamefont {Feng}, \citenamefont {Zhang}, \citenamefont {Smejkal},
  \citenamefont {Sinova}, \citenamefont {Mokrousov},\ and\ \citenamefont
  {Yao}}]{zhou2023crystal}%
  \BibitemOpen
  \bibfield  {author} {\bibinfo {author} {\bibfnamefont {X.}~\bibnamefont
  {Zhou}}, \bibinfo {author} {\bibfnamefont {W.}~\bibnamefont {Feng}}, \bibinfo
  {author} {\bibfnamefont {R.-W.}\ \bibnamefont {Zhang}}, \bibinfo {author}
  {\bibfnamefont {L.}~\bibnamefont {Smejkal}}, \bibinfo {author} {\bibfnamefont
  {J.}~\bibnamefont {Sinova}}, \bibinfo {author} {\bibfnamefont
  {Y.}~\bibnamefont {Mokrousov}}, \ and\ \bibinfo {author} {\bibfnamefont
  {Y.}~\bibnamefont {Yao}},\ }\href@noop {} {\enquote {\bibinfo {title}
  {Crystal thermal transport in altermagnetic ruo2},}\ } (\bibinfo {year}
  {2023}),\ \Eprint {http://arxiv.org/abs/2305.01410} {arXiv:2305.01410
  [cond-mat.mtrl-sci]} \BibitemShut {NoStop}%
\bibitem [{\citenamefont {Liu}\ \emph {et~al.}({\natexlab{a}})\citenamefont
  {Liu}, \citenamefont {Bai}, \citenamefont {Song}, \citenamefont {Ji},
  \citenamefont {Lou}, \citenamefont {Zhang}, \citenamefont {Song},\ and\
  \citenamefont {Jin}}]{liuinverse}%
  \BibitemOpen
  \bibfield  {author} {\bibinfo {author} {\bibfnamefont {Y.}~\bibnamefont
  {Liu}}, \bibinfo {author} {\bibfnamefont {H.}~\bibnamefont {Bai}}, \bibinfo
  {author} {\bibfnamefont {Y.}~\bibnamefont {Song}}, \bibinfo {author}
  {\bibfnamefont {Z.}~\bibnamefont {Ji}}, \bibinfo {author} {\bibfnamefont
  {S.}~\bibnamefont {Lou}}, \bibinfo {author} {\bibfnamefont {Z.}~\bibnamefont
  {Zhang}}, \bibinfo {author} {\bibfnamefont {C.}~\bibnamefont {Song}}, \ and\
  \bibinfo {author} {\bibfnamefont {Q.}~\bibnamefont {Jin}},\ }\href@noop {}
  {\bibfield  {journal} {\bibinfo  {journal} {Advanced Optical Materials}\ ,\
  \bibinfo {pages} {2300177}} ({\natexlab{a}})}\BibitemShut {NoStop}%
\bibitem [{\citenamefont {Qiu}\ \emph {et~al.}(2023)\citenamefont {Qiu},
  \citenamefont {Seifert}, \citenamefont {Huang}, \citenamefont {Zhou},
  \citenamefont {Ka{\v{s}}par}, \citenamefont {Zhang}, \citenamefont {Wu},
  \citenamefont {Fan}, \citenamefont {Zhang}, \citenamefont {Wu} \emph
  {et~al.}}]{qiu2023terahertz}%
  \BibitemOpen
  \bibfield  {author} {\bibinfo {author} {\bibfnamefont {H.}~\bibnamefont
  {Qiu}}, \bibinfo {author} {\bibfnamefont {T.~S.}\ \bibnamefont {Seifert}},
  \bibinfo {author} {\bibfnamefont {L.}~\bibnamefont {Huang}}, \bibinfo
  {author} {\bibfnamefont {Y.}~\bibnamefont {Zhou}}, \bibinfo {author}
  {\bibfnamefont {Z.}~\bibnamefont {Ka{\v{s}}par}}, \bibinfo {author}
  {\bibfnamefont {C.}~\bibnamefont {Zhang}}, \bibinfo {author} {\bibfnamefont
  {J.}~\bibnamefont {Wu}}, \bibinfo {author} {\bibfnamefont {K.}~\bibnamefont
  {Fan}}, \bibinfo {author} {\bibfnamefont {Q.}~\bibnamefont {Zhang}}, \bibinfo
  {author} {\bibfnamefont {D.}~\bibnamefont {Wu}},  \emph {et~al.},\
  }\href@noop {} {\bibfield  {journal} {\bibinfo  {journal} {Advanced Science}\
  ,\ \bibinfo {pages} {2300512}} (\bibinfo {year} {2023})}\BibitemShut
  {NoStop}%
\bibitem [{\citenamefont {Bai}\ \emph {et~al.}(2023)\citenamefont {Bai},
  \citenamefont {Zhang}, \citenamefont {Zhou}, \citenamefont {Chen},
  \citenamefont {Wan}, \citenamefont {Han}, \citenamefont {Zhu}, \citenamefont
  {Liang}, \citenamefont {Su}, \citenamefont {Han} \emph
  {et~al.}}]{bai2023efficient}%
  \BibitemOpen
  \bibfield  {author} {\bibinfo {author} {\bibfnamefont {H.}~\bibnamefont
  {Bai}}, \bibinfo {author} {\bibfnamefont {Y.}~\bibnamefont {Zhang}}, \bibinfo
  {author} {\bibfnamefont {Y.}~\bibnamefont {Zhou}}, \bibinfo {author}
  {\bibfnamefont {P.}~\bibnamefont {Chen}}, \bibinfo {author} {\bibfnamefont
  {C.}~\bibnamefont {Wan}}, \bibinfo {author} {\bibfnamefont {L.}~\bibnamefont
  {Han}}, \bibinfo {author} {\bibfnamefont {W.}~\bibnamefont {Zhu}}, \bibinfo
  {author} {\bibfnamefont {S.}~\bibnamefont {Liang}}, \bibinfo {author}
  {\bibfnamefont {Y.}~\bibnamefont {Su}}, \bibinfo {author} {\bibfnamefont
  {X.}~\bibnamefont {Han}},  \emph {et~al.},\ }\href@noop {} {\bibfield
  {journal} {\bibinfo  {journal} {Physical Review Letters}\ }\textbf {\bibinfo
  {volume} {130}},\ \bibinfo {pages} {216701} (\bibinfo {year}
  {2023})}\BibitemShut {NoStop}%
\bibitem [{\citenamefont {Ouassou}\ \emph {et~al.}(2023)\citenamefont
  {Ouassou}, \citenamefont {Brataas},\ and\ \citenamefont
  {Linder}}]{Ouassou23}%
  \BibitemOpen
  \bibfield  {author} {\bibinfo {author} {\bibfnamefont {J.~A.}\ \bibnamefont
  {Ouassou}}, \bibinfo {author} {\bibfnamefont {A.}~\bibnamefont {Brataas}}, \
  and\ \bibinfo {author} {\bibfnamefont {J.}~\bibnamefont {Linder}},\
  }\href@noop {} {\enquote {\bibinfo {title} {Josephson effect in
  altermagnets},}\ } (\bibinfo {year} {2023}),\ \Eprint
  {http://arxiv.org/abs/2301.03603} {arXiv:2301.03603 [cond-mat.supr-con]}
  \BibitemShut {NoStop}%
\bibitem [{\citenamefont {Shao}\ \emph {et~al.}(2021)\citenamefont {Shao},
  \citenamefont {Zhang}, \citenamefont {Li}, \citenamefont {Eom},\ and\
  \citenamefont {Tsymbal}}]{Shao21}%
  \BibitemOpen
  \bibfield  {author} {\bibinfo {author} {\bibfnamefont {D.-F.}\ \bibnamefont
  {Shao}}, \bibinfo {author} {\bibfnamefont {S.-H.}\ \bibnamefont {Zhang}},
  \bibinfo {author} {\bibfnamefont {M.}~\bibnamefont {Li}}, \bibinfo {author}
  {\bibfnamefont {C.-B.}\ \bibnamefont {Eom}}, \ and\ \bibinfo {author}
  {\bibfnamefont {E.~Y.}\ \bibnamefont {Tsymbal}},\ }\href {\doibase
  10.1038/s41467-021-26915-3} {\bibfield  {journal} {\bibinfo  {journal}
  {Nature Communications}\ }\textbf {\bibinfo {volume} {12}} (\bibinfo {year}
  {2021}),\ 10.1038/s41467-021-26915-3}\BibitemShut {NoStop}%
\bibitem [{\citenamefont {Rimmler}\ \emph {et~al.}(2023)\citenamefont
  {Rimmler}, \citenamefont {Hazra}, \citenamefont {Pal}, \citenamefont
  {Mohseni}, \citenamefont {Taylor}, \citenamefont {Bedoya-Pinto},
  \citenamefont {Deniz}, \citenamefont {Tangi}, \citenamefont {Kostanovskiy},
  \citenamefont {Luo} \emph {et~al.}}]{rimmler2023atomic}%
  \BibitemOpen
  \bibfield  {author} {\bibinfo {author} {\bibfnamefont {B.~H.}\ \bibnamefont
  {Rimmler}}, \bibinfo {author} {\bibfnamefont {B.~K.}\ \bibnamefont {Hazra}},
  \bibinfo {author} {\bibfnamefont {B.}~\bibnamefont {Pal}}, \bibinfo {author}
  {\bibfnamefont {K.}~\bibnamefont {Mohseni}}, \bibinfo {author} {\bibfnamefont
  {J.~M.}\ \bibnamefont {Taylor}}, \bibinfo {author} {\bibfnamefont
  {A.}~\bibnamefont {Bedoya-Pinto}}, \bibinfo {author} {\bibfnamefont
  {H.}~\bibnamefont {Deniz}}, \bibinfo {author} {\bibfnamefont {M.~R.}\
  \bibnamefont {Tangi}}, \bibinfo {author} {\bibfnamefont {I.}~\bibnamefont
  {Kostanovskiy}}, \bibinfo {author} {\bibfnamefont {C.}~\bibnamefont {Luo}},
  \emph {et~al.},\ }\href@noop {} {\bibfield  {journal} {\bibinfo  {journal}
  {Advanced Materials}\ ,\ \bibinfo {pages} {2209616}} (\bibinfo {year}
  {2023})}\BibitemShut {NoStop}%
\bibitem [{\citenamefont {Ahn}\ and\ \citenamefont
  {Zhao}(2023)}]{ahn2023flipping}%
  \BibitemOpen
  \bibfield  {author} {\bibinfo {author} {\bibfnamefont {Y.}~\bibnamefont
  {Ahn}}\ and\ \bibinfo {author} {\bibfnamefont {L.}~\bibnamefont {Zhao}},\
  }\href@noop {} {\bibfield  {journal} {\bibinfo  {journal} {Nature Materials}\
  }\textbf {\bibinfo {volume} {22}},\ \bibinfo {pages} {536} (\bibinfo {year}
  {2023})}\BibitemShut {NoStop}%
\bibitem [{\citenamefont {Cuono}\ \emph
  {et~al.}(2023{\natexlab{a}})\citenamefont {Cuono}, \citenamefont {Sattigeri},
  \citenamefont {Skolimowski},\ and\ \citenamefont {Autieri}}]{Cuono23orbital}%
  \BibitemOpen
  \bibfield  {author} {\bibinfo {author} {\bibfnamefont {G.}~\bibnamefont
  {Cuono}}, \bibinfo {author} {\bibfnamefont {R.~M.}\ \bibnamefont
  {Sattigeri}}, \bibinfo {author} {\bibfnamefont {J.}~\bibnamefont
  {Skolimowski}}, \ and\ \bibinfo {author} {\bibfnamefont {C.}~\bibnamefont
  {Autieri}},\ }\href@noop {} {\enquote {\bibinfo {title} {Orbital-selective
  altermagnetism and correlation-enhanced spin-splitting in transition metal
  oxides},}\ } (\bibinfo {year} {2023}{\natexlab{a}}),\ \Eprint
  {http://arxiv.org/abs/2306.17497} {arXiv:2306.17497 [cond-mat.str-el]}
  \BibitemShut {NoStop}%
\bibitem [{\citenamefont {Nguyen}\ and\ \citenamefont
  {Yamauchi}(2023)}]{PhysRevB.107.155126}%
  \BibitemOpen
  \bibfield  {author} {\bibinfo {author} {\bibfnamefont {T.~P.~T.}\
  \bibnamefont {Nguyen}}\ and\ \bibinfo {author} {\bibfnamefont
  {K.}~\bibnamefont {Yamauchi}},\ }\href {\doibase 10.1103/PhysRevB.107.155126}
  {\bibfield  {journal} {\bibinfo  {journal} {Phys. Rev. B}\ }\textbf {\bibinfo
  {volume} {107}},\ \bibinfo {pages} {155126} (\bibinfo {year}
  {2023})}\BibitemShut {NoStop}%
\bibitem [{\citenamefont {Autieri}\ \emph {et~al.}(2023)\citenamefont
  {Autieri}, \citenamefont {Cuoco}, \citenamefont {Cuono}, \citenamefont
  {Picozzi},\ and\ \citenamefont {Noce}}]{AUTIERI2023414407}%
  \BibitemOpen
  \bibfield  {author} {\bibinfo {author} {\bibfnamefont {C.}~\bibnamefont
  {Autieri}}, \bibinfo {author} {\bibfnamefont {M.}~\bibnamefont {Cuoco}},
  \bibinfo {author} {\bibfnamefont {G.}~\bibnamefont {Cuono}}, \bibinfo
  {author} {\bibfnamefont {S.}~\bibnamefont {Picozzi}}, \ and\ \bibinfo
  {author} {\bibfnamefont {C.}~\bibnamefont {Noce}},\ }\href {\doibase
  10.1016/j.physb.2022.414407} {\bibfield  {journal} {\bibinfo  {journal}
  {Physica B: Condensed Matter}\ }\textbf {\bibinfo {volume} {648}},\ \bibinfo
  {pages} {414407} (\bibinfo {year} {2023})}\BibitemShut {NoStop}%
\bibitem [{\citenamefont {Kriegner}\ \emph {et~al.}(2017)\citenamefont
  {Kriegner}, \citenamefont {Reichlova}, \citenamefont {Grenzer}, \citenamefont
  {Schmidt}, \citenamefont {Ressouche}, \citenamefont {Godinho}, \citenamefont
  {Wagner}, \citenamefont {Martin}, \citenamefont {Shick}, \citenamefont
  {Volobuev}, \citenamefont {Springholz}, \citenamefont {Hol\'y}, \citenamefont
  {Wunderlich}, \citenamefont {Jungwirth},\ and\ \citenamefont
  {V\'yborn\'y}}]{PhysRevB.96.214418}%
  \BibitemOpen
  \bibfield  {author} {\bibinfo {author} {\bibfnamefont {D.}~\bibnamefont
  {Kriegner}}, \bibinfo {author} {\bibfnamefont {H.}~\bibnamefont {Reichlova}},
  \bibinfo {author} {\bibfnamefont {J.}~\bibnamefont {Grenzer}}, \bibinfo
  {author} {\bibfnamefont {W.}~\bibnamefont {Schmidt}}, \bibinfo {author}
  {\bibfnamefont {E.}~\bibnamefont {Ressouche}}, \bibinfo {author}
  {\bibfnamefont {J.}~\bibnamefont {Godinho}}, \bibinfo {author} {\bibfnamefont
  {T.}~\bibnamefont {Wagner}}, \bibinfo {author} {\bibfnamefont {S.~Y.}\
  \bibnamefont {Martin}}, \bibinfo {author} {\bibfnamefont {A.~B.}\
  \bibnamefont {Shick}}, \bibinfo {author} {\bibfnamefont {V.~V.}\ \bibnamefont
  {Volobuev}}, \bibinfo {author} {\bibfnamefont {G.}~\bibnamefont
  {Springholz}}, \bibinfo {author} {\bibfnamefont {V.}~\bibnamefont {Hol\'y}},
  \bibinfo {author} {\bibfnamefont {J.}~\bibnamefont {Wunderlich}}, \bibinfo
  {author} {\bibfnamefont {T.}~\bibnamefont {Jungwirth}}, \ and\ \bibinfo
  {author} {\bibfnamefont {K.}~\bibnamefont {V\'yborn\'y}},\ }\href {\doibase
  10.1103/PhysRevB.96.214418} {\bibfield  {journal} {\bibinfo  {journal} {Phys.
  Rev. B}\ }\textbf {\bibinfo {volume} {96}},\ \bibinfo {pages} {214418}
  (\bibinfo {year} {2017})}\BibitemShut {NoStop}%
\bibitem [{\citenamefont {Bossini}\ \emph {et~al.}(2020)\citenamefont
  {Bossini}, \citenamefont {Terschanski}, \citenamefont {Mertens},
  \citenamefont {Springholz}, \citenamefont {Bonanni}, \citenamefont {Uhrig},\
  and\ \citenamefont {Cinchetti}}]{Bossini_2020}%
  \BibitemOpen
  \bibfield  {author} {\bibinfo {author} {\bibfnamefont {D.}~\bibnamefont
  {Bossini}}, \bibinfo {author} {\bibfnamefont {M.}~\bibnamefont
  {Terschanski}}, \bibinfo {author} {\bibfnamefont {F.}~\bibnamefont
  {Mertens}}, \bibinfo {author} {\bibfnamefont {G.}~\bibnamefont {Springholz}},
  \bibinfo {author} {\bibfnamefont {A.}~\bibnamefont {Bonanni}}, \bibinfo
  {author} {\bibfnamefont {G.~S.}\ \bibnamefont {Uhrig}}, \ and\ \bibinfo
  {author} {\bibfnamefont {M.}~\bibnamefont {Cinchetti}},\ }\href {\doibase
  10.1088/1367-2630/aba0e7} {\bibfield  {journal} {\bibinfo  {journal} {New J.
  Phys.}\ }\textbf {\bibinfo {volume} {22}},\ \bibinfo {pages} {083029}
  (\bibinfo {year} {2020})}\BibitemShut {NoStop}%
\bibitem [{\citenamefont {Mazin}(2023)}]{Mazin23}%
  \BibitemOpen
  \bibfield  {author} {\bibinfo {author} {\bibfnamefont {I.~I.}\ \bibnamefont
  {Mazin}},\ }\href {\doibase 10.1103/PhysRevB.107.L100418} {\bibfield
  {journal} {\bibinfo  {journal} {Phys. Rev. B}\ }\textbf {\bibinfo {volume}
  {107}},\ \bibinfo {pages} {L100418} (\bibinfo {year} {2023})}\BibitemShut
  {NoStop}%
\bibitem [{\citenamefont {Aoyama}\ and\ \citenamefont
  {Ohgushi}(2023)}]{Aoyama23}%
  \BibitemOpen
  \bibfield  {author} {\bibinfo {author} {\bibfnamefont {T.}~\bibnamefont
  {Aoyama}}\ and\ \bibinfo {author} {\bibfnamefont {K.}~\bibnamefont
  {Ohgushi}},\ }\href@noop {} {\enquote {\bibinfo {title} {Piezomagnetic
  properties in altermagnetic mnte},}\ } (\bibinfo {year} {2023}),\ \Eprint
  {http://arxiv.org/abs/2305.14786} {arXiv:2305.14786 [cond-mat.mtrl-sci]}
  \BibitemShut {NoStop}%
\bibitem [{\citenamefont {Hariki}\ \emph {et~al.}(2023)\citenamefont {Hariki},
  \citenamefont {Yamaguchi}, \citenamefont {Kriegner}, \citenamefont {Edmonds},
  \citenamefont {Wadley}, \citenamefont {Dhesi}, \citenamefont {Springholz},
  \citenamefont {Šmejkal}, \citenamefont {Výborný}, \citenamefont
  {Jungwirth},\ and\ \citenamefont {Kuneš}}]{Hariki23}%
  \BibitemOpen
  \bibfield  {author} {\bibinfo {author} {\bibfnamefont {A.}~\bibnamefont
  {Hariki}}, \bibinfo {author} {\bibfnamefont {T.}~\bibnamefont {Yamaguchi}},
  \bibinfo {author} {\bibfnamefont {D.}~\bibnamefont {Kriegner}}, \bibinfo
  {author} {\bibfnamefont {K.~W.}\ \bibnamefont {Edmonds}}, \bibinfo {author}
  {\bibfnamefont {P.}~\bibnamefont {Wadley}}, \bibinfo {author} {\bibfnamefont
  {S.~S.}\ \bibnamefont {Dhesi}}, \bibinfo {author} {\bibfnamefont
  {G.}~\bibnamefont {Springholz}}, \bibinfo {author} {\bibfnamefont
  {L.}~\bibnamefont {Šmejkal}}, \bibinfo {author} {\bibfnamefont
  {K.}~\bibnamefont {Výborný}}, \bibinfo {author} {\bibfnamefont
  {T.}~\bibnamefont {Jungwirth}}, \ and\ \bibinfo {author} {\bibfnamefont
  {J.}~\bibnamefont {Kuneš}},\ }\href@noop {} {\enquote {\bibinfo {title}
  {X-ray magnetic circular dichroism in altermagnetic $\alpha$-mnte},}\ }
  (\bibinfo {year} {2023}),\ \Eprint {http://arxiv.org/abs/2305.03588}
  {arXiv:2305.03588 [cond-mat.mtrl-sci]} \BibitemShut {NoStop}%
\bibitem [{\citenamefont {Junior}\ \emph {et~al.}(2023)\citenamefont {Junior},
  \citenamefont {de~Mare}, \citenamefont {Zollner}, \citenamefont {Ahn},
  \citenamefont {Erlingsson}, \citenamefont {van Schilfgaarde},\ and\
  \citenamefont {V{\`y}born{\`y}}}]{junior2023sensitivity}%
  \BibitemOpen
  \bibfield  {author} {\bibinfo {author} {\bibfnamefont {P.~E.~F.}\
  \bibnamefont {Junior}}, \bibinfo {author} {\bibfnamefont {K.~A.}\
  \bibnamefont {de~Mare}}, \bibinfo {author} {\bibfnamefont {K.}~\bibnamefont
  {Zollner}}, \bibinfo {author} {\bibfnamefont {K.-h.}\ \bibnamefont {Ahn}},
  \bibinfo {author} {\bibfnamefont {S.~I.}\ \bibnamefont {Erlingsson}},
  \bibinfo {author} {\bibfnamefont {M.}~\bibnamefont {van Schilfgaarde}}, \
  and\ \bibinfo {author} {\bibfnamefont {K.}~\bibnamefont {V{\`y}born{\`y}}},\
  }\href@noop {} {\bibfield  {journal} {\bibinfo  {journal} {Physical Review
  B}\ }\textbf {\bibinfo {volume} {107}},\ \bibinfo {pages} {L100417} (\bibinfo
  {year} {2023})}\BibitemShut {NoStop}%
\bibitem [{\citenamefont {Xiao}\ \emph {et~al.}(2018)\citenamefont {Xiao},
  \citenamefont {Jiang}, \citenamefont {Shin}, \citenamefont {Wang},
  \citenamefont {Wang}, \citenamefont {Zhao}, \citenamefont {Liu},
  \citenamefont {Wu}, \citenamefont {Chan}, \citenamefont {Samarth},\ and\
  \citenamefont {Chang}}]{PhysRevLett.120.056801}%
  \BibitemOpen
  \bibfield  {author} {\bibinfo {author} {\bibfnamefont {D.}~\bibnamefont
  {Xiao}}, \bibinfo {author} {\bibfnamefont {J.}~\bibnamefont {Jiang}},
  \bibinfo {author} {\bibfnamefont {J.-H.}\ \bibnamefont {Shin}}, \bibinfo
  {author} {\bibfnamefont {W.}~\bibnamefont {Wang}}, \bibinfo {author}
  {\bibfnamefont {F.}~\bibnamefont {Wang}}, \bibinfo {author} {\bibfnamefont
  {Y.-F.}\ \bibnamefont {Zhao}}, \bibinfo {author} {\bibfnamefont
  {C.}~\bibnamefont {Liu}}, \bibinfo {author} {\bibfnamefont {W.}~\bibnamefont
  {Wu}}, \bibinfo {author} {\bibfnamefont {M.~H.~W.}\ \bibnamefont {Chan}},
  \bibinfo {author} {\bibfnamefont {N.}~\bibnamefont {Samarth}}, \ and\
  \bibinfo {author} {\bibfnamefont {C.-Z.}\ \bibnamefont {Chang}},\ }\href
  {\doibase 10.1103/PhysRevLett.120.056801} {\bibfield  {journal} {\bibinfo
  {journal} {Phys. Rev. Lett.}\ }\textbf {\bibinfo {volume} {120}},\ \bibinfo
  {pages} {056801} (\bibinfo {year} {2018})}\BibitemShut {NoStop}%
\bibitem [{\citenamefont {Lei}\ \emph {et~al.}(2021)\citenamefont {Lei},
  \citenamefont {Chen},\ and\ \citenamefont {MacDonald}}]{lei2021large}%
  \BibitemOpen
  \bibfield  {author} {\bibinfo {author} {\bibfnamefont {C.}~\bibnamefont
  {Lei}}, \bibinfo {author} {\bibfnamefont {H.}~\bibnamefont {Chen}}, \ and\
  \bibinfo {author} {\bibfnamefont {A.~H.}\ \bibnamefont {MacDonald}},\
  }\href@noop {} {\enquote {\bibinfo {title} {Large anomalous hall effect in
  topological insulators proximitized by collinear antiferromagnets},}\ }
  (\bibinfo {year} {2021}),\ \Eprint {http://arxiv.org/abs/2107.02307}
  {arXiv:2107.02307 [cond-mat.mes-hall]} \BibitemShut {NoStop}%
\bibitem [{\citenamefont {Pournaghavi}\ \emph {et~al.}(2021)\citenamefont
  {Pournaghavi}, \citenamefont {Islam}, \citenamefont {Islam}, \citenamefont
  {Autieri}, \citenamefont {Dietl},\ and\ \citenamefont
  {Canali}}]{PhysRevB.103.195308}%
  \BibitemOpen
  \bibfield  {author} {\bibinfo {author} {\bibfnamefont {N.}~\bibnamefont
  {Pournaghavi}}, \bibinfo {author} {\bibfnamefont {M.~F.}\ \bibnamefont
  {Islam}}, \bibinfo {author} {\bibfnamefont {R.}~\bibnamefont {Islam}},
  \bibinfo {author} {\bibfnamefont {C.}~\bibnamefont {Autieri}}, \bibinfo
  {author} {\bibfnamefont {T.}~\bibnamefont {Dietl}}, \ and\ \bibinfo {author}
  {\bibfnamefont {C.~M.}\ \bibnamefont {Canali}},\ }\href {\doibase
  10.1103/PhysRevB.103.195308} {\bibfield  {journal} {\bibinfo  {journal}
  {Phys. Rev. B}\ }\textbf {\bibinfo {volume} {103}},\ \bibinfo {pages}
  {195308} (\bibinfo {year} {2021})}\BibitemShut {NoStop}%
\bibitem [{\citenamefont {Guo}\ \emph {et~al.}(2023{\natexlab{b}})\citenamefont
  {Guo}, \citenamefont {Guo}, \citenamefont {Cheng}, \citenamefont {Wang},\
  and\ \citenamefont {Ang}}]{guo2023piezoelectric}%
  \BibitemOpen
  \bibfield  {author} {\bibinfo {author} {\bibfnamefont {S.-D.}\ \bibnamefont
  {Guo}}, \bibinfo {author} {\bibfnamefont {X.-S.}\ \bibnamefont {Guo}},
  \bibinfo {author} {\bibfnamefont {K.}~\bibnamefont {Cheng}}, \bibinfo
  {author} {\bibfnamefont {K.}~\bibnamefont {Wang}}, \ and\ \bibinfo {author}
  {\bibfnamefont {Y.~S.}\ \bibnamefont {Ang}},\ }\href@noop {} {\enquote
  {\bibinfo {title} {Piezoelectric altermagnetism and spin-valley polarization
  in janus monolayer $\mathrm{Cr_2SO}$},}\ } (\bibinfo {year}
  {2023}{\natexlab{b}}),\ \Eprint {http://arxiv.org/abs/2306.04094}
  {arXiv:2306.04094 [cond-mat.mtrl-sci]} \BibitemShut {NoStop}%
\bibitem [{\citenamefont {Liu}\ \emph {et~al.}({\natexlab{b}})\citenamefont
  {Liu}, \citenamefont {Bai}, \citenamefont {Song}, \citenamefont {Ji},
  \citenamefont {Lou}, \citenamefont {Zhang}, \citenamefont {Song},\ and\
  \citenamefont {Jin}}]{https://doi.org/10.1002/adom.202300177}%
  \BibitemOpen
  \bibfield  {author} {\bibinfo {author} {\bibfnamefont {Y.}~\bibnamefont
  {Liu}}, \bibinfo {author} {\bibfnamefont {H.}~\bibnamefont {Bai}}, \bibinfo
  {author} {\bibfnamefont {Y.}~\bibnamefont {Song}}, \bibinfo {author}
  {\bibfnamefont {Z.}~\bibnamefont {Ji}}, \bibinfo {author} {\bibfnamefont
  {S.}~\bibnamefont {Lou}}, \bibinfo {author} {\bibfnamefont {Z.}~\bibnamefont
  {Zhang}}, \bibinfo {author} {\bibfnamefont {C.}~\bibnamefont {Song}}, \ and\
  \bibinfo {author} {\bibfnamefont {Q.}~\bibnamefont {Jin}},\ }\href {\doibase
  https://doi.org/10.1002/adom.202300177} {\bibfield  {journal} {\bibinfo
  {journal} {Advanced Optical Materials}\ }\textbf {\bibinfo {volume} {n/a}},\
  \bibinfo {pages} {2300177} ({\natexlab{b}})}\BibitemShut {NoStop}%
\bibitem [{\citenamefont {Keshavarz}\ \emph {et~al.}(2017)\citenamefont
  {Keshavarz}, \citenamefont {Kvashnin}, \citenamefont {Rodrigues},
  \citenamefont {Pereiro}, \citenamefont {Di~Marco}, \citenamefont {Autieri},
  \citenamefont {Nordstr\"om}, \citenamefont {Solovyev}, \citenamefont
  {Sanyal},\ and\ \citenamefont {Eriksson}}]{PhysRevB.95.115120}%
  \BibitemOpen
  \bibfield  {author} {\bibinfo {author} {\bibfnamefont {S.}~\bibnamefont
  {Keshavarz}}, \bibinfo {author} {\bibfnamefont {Y.~O.}\ \bibnamefont
  {Kvashnin}}, \bibinfo {author} {\bibfnamefont {D.~C.~M.}\ \bibnamefont
  {Rodrigues}}, \bibinfo {author} {\bibfnamefont {M.}~\bibnamefont {Pereiro}},
  \bibinfo {author} {\bibfnamefont {I.}~\bibnamefont {Di~Marco}}, \bibinfo
  {author} {\bibfnamefont {C.}~\bibnamefont {Autieri}}, \bibinfo {author}
  {\bibfnamefont {L.}~\bibnamefont {Nordstr\"om}}, \bibinfo {author}
  {\bibfnamefont {I.~V.}\ \bibnamefont {Solovyev}}, \bibinfo {author}
  {\bibfnamefont {B.}~\bibnamefont {Sanyal}}, \ and\ \bibinfo {author}
  {\bibfnamefont {O.}~\bibnamefont {Eriksson}},\ }\href {\doibase
  10.1103/PhysRevB.95.115120} {\bibfield  {journal} {\bibinfo  {journal} {Phys.
  Rev. B}\ }\textbf {\bibinfo {volume} {95}},\ \bibinfo {pages} {115120}
  (\bibinfo {year} {2017})}\BibitemShut {NoStop}%
\bibitem [{\citenamefont {Cuono}\ \emph
  {et~al.}(2023{\natexlab{b}})\citenamefont {Cuono}, \citenamefont {Sattigeri},
  \citenamefont {Autieri},\ and\ \citenamefont {Dietl}}]{Cuono23EuCd2As2}%
  \BibitemOpen
  \bibfield  {author} {\bibinfo {author} {\bibfnamefont {G.}~\bibnamefont
  {Cuono}}, \bibinfo {author} {\bibfnamefont {R.~M.}\ \bibnamefont
  {Sattigeri}}, \bibinfo {author} {\bibfnamefont {C.}~\bibnamefont {Autieri}},
  \ and\ \bibinfo {author} {\bibfnamefont {T.}~\bibnamefont {Dietl}},\
  }\href@noop {} {\enquote {\bibinfo {title} {Ab-initio overestimation of the
  topological region in eu-based compounds},}\ } (\bibinfo {year}
  {2023}{\natexlab{b}}),\ \Eprint {http://arxiv.org/abs/2305.10804}
  {arXiv:2305.10804 [cond-mat.str-el]} \BibitemShut {NoStop}%
\bibitem [{\citenamefont {Jain}\ \emph {et~al.}(2013)\citenamefont {Jain},
  \citenamefont {Ong}, \citenamefont {Hautier}, \citenamefont {Chen},
  \citenamefont {Richards}, \citenamefont {Dacek}, \citenamefont {Cholia},
  \citenamefont {Gunter}, \citenamefont {Skinner}, \citenamefont {Ceder} \emph
  {et~al.}}]{materials-project}%
  \BibitemOpen
  \bibfield  {author} {\bibinfo {author} {\bibfnamefont {A.}~\bibnamefont
  {Jain}}, \bibinfo {author} {\bibfnamefont {S.~P.}\ \bibnamefont {Ong}},
  \bibinfo {author} {\bibfnamefont {G.}~\bibnamefont {Hautier}}, \bibinfo
  {author} {\bibfnamefont {W.}~\bibnamefont {Chen}}, \bibinfo {author}
  {\bibfnamefont {W.~D.}\ \bibnamefont {Richards}}, \bibinfo {author}
  {\bibfnamefont {S.}~\bibnamefont {Dacek}}, \bibinfo {author} {\bibfnamefont
  {S.}~\bibnamefont {Cholia}}, \bibinfo {author} {\bibfnamefont
  {D.}~\bibnamefont {Gunter}}, \bibinfo {author} {\bibfnamefont
  {D.}~\bibnamefont {Skinner}}, \bibinfo {author} {\bibfnamefont
  {G.}~\bibnamefont {Ceder}},  \emph {et~al.},\ }\href@noop {} {\bibfield
  {journal} {\bibinfo  {journal} {APL materials}\ }\textbf {\bibinfo {volume}
  {1}},\ \bibinfo {pages} {011002} (\bibinfo {year} {2013})}\BibitemShut
  {NoStop}%
\bibitem [{\citenamefont {Giannozzi}\ \emph {et~al.}(2009)\citenamefont
  {Giannozzi}, \citenamefont {Baroni}, \citenamefont {Bonini}, \citenamefont
  {Calandra}, \citenamefont {Car}, \citenamefont {Cavazzoni}, \citenamefont
  {Ceresoli}, \citenamefont {Chiarotti}, \citenamefont {Cococcioni},
  \citenamefont {Dabo} \emph {et~al.}}]{giannozzi2009quantum}%
  \BibitemOpen
  \bibfield  {author} {\bibinfo {author} {\bibfnamefont {P.}~\bibnamefont
  {Giannozzi}}, \bibinfo {author} {\bibfnamefont {S.}~\bibnamefont {Baroni}},
  \bibinfo {author} {\bibfnamefont {N.}~\bibnamefont {Bonini}}, \bibinfo
  {author} {\bibfnamefont {M.}~\bibnamefont {Calandra}}, \bibinfo {author}
  {\bibfnamefont {R.}~\bibnamefont {Car}}, \bibinfo {author} {\bibfnamefont
  {C.}~\bibnamefont {Cavazzoni}}, \bibinfo {author} {\bibfnamefont
  {D.}~\bibnamefont {Ceresoli}}, \bibinfo {author} {\bibfnamefont {G.~L.}\
  \bibnamefont {Chiarotti}}, \bibinfo {author} {\bibfnamefont {M.}~\bibnamefont
  {Cococcioni}}, \bibinfo {author} {\bibfnamefont {I.}~\bibnamefont {Dabo}},
  \emph {et~al.},\ }\href@noop {} {\bibfield  {journal} {\bibinfo  {journal}
  {Journal of physics: Condensed matter}\ }\textbf {\bibinfo {volume} {21}},\
  \bibinfo {pages} {395502} (\bibinfo {year} {2009})}\BibitemShut {NoStop}%
\bibitem [{\citenamefont {Prandini}\ \emph {et~al.}(2018)\citenamefont
  {Prandini}, \citenamefont {Marrazzo}, \citenamefont {Castelli}, \citenamefont
  {Mounet},\ and\ \citenamefont {Marzari}}]{pseudos}%
  \BibitemOpen
  \bibfield  {author} {\bibinfo {author} {\bibfnamefont {G.}~\bibnamefont
  {Prandini}}, \bibinfo {author} {\bibfnamefont {A.}~\bibnamefont {Marrazzo}},
  \bibinfo {author} {\bibfnamefont {I.~E.}\ \bibnamefont {Castelli}}, \bibinfo
  {author} {\bibfnamefont {N.}~\bibnamefont {Mounet}}, \ and\ \bibinfo {author}
  {\bibfnamefont {N.}~\bibnamefont {Marzari}},\ }\href@noop {} {\bibfield
  {journal} {\bibinfo  {journal} {npj Computational Materials}\ }\textbf
  {\bibinfo {volume} {4}},\ \bibinfo {pages} {72} (\bibinfo {year}
  {2018})}\BibitemShut {NoStop}%
\bibitem [{\citenamefont {Perdew}\ \emph {et~al.}(1996)\citenamefont {Perdew},
  \citenamefont {Burke},\ and\ \citenamefont
  {Ernzerhof}}]{perdew1996generalized}%
  \BibitemOpen
  \bibfield  {author} {\bibinfo {author} {\bibfnamefont {J.~P.}\ \bibnamefont
  {Perdew}}, \bibinfo {author} {\bibfnamefont {K.}~\bibnamefont {Burke}}, \
  and\ \bibinfo {author} {\bibfnamefont {M.}~\bibnamefont {Ernzerhof}},\
  }\href@noop {} {\bibfield  {journal} {\bibinfo  {journal} {Physical review
  letters}\ }\textbf {\bibinfo {volume} {77}},\ \bibinfo {pages} {3865}
  (\bibinfo {year} {1996})}\BibitemShut {NoStop}%
\bibitem [{\citenamefont {Monkhorst}\ and\ \citenamefont
  {Pack}(1976)}]{monkhorst1976special}%
  \BibitemOpen
  \bibfield  {author} {\bibinfo {author} {\bibfnamefont {H.~J.}\ \bibnamefont
  {Monkhorst}}\ and\ \bibinfo {author} {\bibfnamefont {J.~D.}\ \bibnamefont
  {Pack}},\ }\href@noop {} {\bibfield  {journal} {\bibinfo  {journal} {Physical
  review B}\ }\textbf {\bibinfo {volume} {13}},\ \bibinfo {pages} {5188}
  (\bibinfo {year} {1976})}\BibitemShut {NoStop}%
\bibitem [{\citenamefont {Mostofi}\ \emph {et~al.}(2014)\citenamefont
  {Mostofi}, \citenamefont {Yates}, \citenamefont {Pizzi}, \citenamefont {Lee},
  \citenamefont {Souza}, \citenamefont {Vanderbilt},\ and\ \citenamefont
  {Marzari}}]{w90}%
  \BibitemOpen
  \bibfield  {author} {\bibinfo {author} {\bibfnamefont {A.~A.}\ \bibnamefont
  {Mostofi}}, \bibinfo {author} {\bibfnamefont {J.~R.}\ \bibnamefont {Yates}},
  \bibinfo {author} {\bibfnamefont {G.}~\bibnamefont {Pizzi}}, \bibinfo
  {author} {\bibfnamefont {Y.-S.}\ \bibnamefont {Lee}}, \bibinfo {author}
  {\bibfnamefont {I.}~\bibnamefont {Souza}}, \bibinfo {author} {\bibfnamefont
  {D.}~\bibnamefont {Vanderbilt}}, \ and\ \bibinfo {author} {\bibfnamefont
  {N.}~\bibnamefont {Marzari}},\ }\href@noop {} {\bibfield  {journal} {\bibinfo
   {journal} {Computer Physics Communications}\ }\textbf {\bibinfo {volume}
  {185}},\ \bibinfo {pages} {2309} (\bibinfo {year} {2014})}\BibitemShut
  {NoStop}%
\bibitem [{\citenamefont {Wu}\ \emph {et~al.}(2018)\citenamefont {Wu},
  \citenamefont {Zhang}, \citenamefont {Song}, \citenamefont {Troyer},\ and\
  \citenamefont {Soluyanov}}]{wanniertools}%
  \BibitemOpen
  \bibfield  {author} {\bibinfo {author} {\bibfnamefont {Q.}~\bibnamefont
  {Wu}}, \bibinfo {author} {\bibfnamefont {S.}~\bibnamefont {Zhang}}, \bibinfo
  {author} {\bibfnamefont {H.-F.}\ \bibnamefont {Song}}, \bibinfo {author}
  {\bibfnamefont {M.}~\bibnamefont {Troyer}}, \ and\ \bibinfo {author}
  {\bibfnamefont {A.~A.}\ \bibnamefont {Soluyanov}},\ }\href@noop {} {\bibfield
   {journal} {\bibinfo  {journal} {Computer Physics Communications}\ }\textbf
  {\bibinfo {volume} {224}},\ \bibinfo {pages} {405} (\bibinfo {year}
  {2018})}\BibitemShut {NoStop}%
\end{thebibliography}%
\end{document}